\documentclass[a4paper,usenatbib]{mnras}

\makeatletter

\newcommand{\Rmnum}[1]{\expandafter\@slowromancap\romannumeral #1@}
\newcommand{\lsim}{\lesssim}
\newcommand{\gsim}{\lower0.6ex\vbox{\hbox{$\buildrel{\textstyle >}\over{\sim}\ $}}}

\makeatother

\usepackage{newtxtext,newtxmath}
\usepackage{graphicx}	
\usepackage{amsmath}	
\usepackage{subcaption}
\usepackage{booktabs}
\usepackage[dvipsnames,svgnames]{xcolor}
\usepackage{color}
\newcommand{\yy}[1]{\textcolor{black}{#1}}

\def\asterix{\texttt{Astrid} }
\def\bluetides{\texttt{BlueTides}}

\def\mbh{\, M_{\rm BH}}
\def\msd{\, M_{\rm sd}}
\def\mstar{\, M_*}

\def\kpc{\, {\rm kpc}}
\def\hmpc{h^{-1}{\rm Mpc}}

\def\hkpc{h^{-1}\, {\rm kpc}}

\def\hmsun{{h^{-1} M_{\odot}}}
\def\msun{\, M_{\odot}}
\def\Lx{L_{\rm X}}
\def\NH{N_{\rm H}}

\title[ASTRID BHs]{The ASTRID simulation: the evolution of Supermassive Black Holes}

\author[Y.~Ni et al.]{
Yueying Ni,$^{1,2}$\thanks{E-mail: yueyingn@andrew.cmu.edu}
Tiziana Di Matteo,$^{1,2}$
Simeon Bird,$^{3}$
Rupert Croft,$^{1,2}$
Yu Feng,$^{4}$
\newauthor
Nianyi Chen,$^{1}$
Michael Tremmel,$^{5}$
Colin DeGraf,$^{6}$
Yin Li,$^{7}$
\\
$^{1}$ McWilliams Center for Cosmology, Department of Physics, Carnegie Mellon University, Pittsburgh, PA 15213 \\
$^{2}$ NSF AI Planning Institute for Physics of the Future, 
Carnegie   Mellon  University, Pittsburgh, PA 15213, USA \\
$^{3}$ Department of Physics \& Astronomy, University of California, Riverside, 900 University Ave., Riverside, CA 92521, USA\\
$^{4}$ Berkeley Center for Cosmological Physics and Department of Physics, University of California, Berkeley, CA 94720, USA \\
$^{5}$ Astronomy Department, Yale University, P.O. Box 208120, New Haven, CT 06520, USA \\
$^{6}$ Department of Physics, Truman State University, Kirksville, MO 63501, USA \\
$^{7}$ Center for Computational Astrophysics \& Center for Computational Mathematics, Flatiron Institute, 162 5th Avenue, New York, NY 10010 \\
}

\date{Accepted XXX. Received YYY; in original form ZZZ}

\pubyear{2021}
\begin{document}
\maketitle

\begin{abstract}
We present the evolution of black holes (BHs) and their relationship with their host galaxies in \asterix, a large-volume cosmological hydrodynamical simulation with box size 250 $\hmpc$ containing $2\times5500^3$ particles evolved to $z=3$.
\asterix statistically models BH gas accretion and AGN feedback to their environments, applies a power-law distribution for BH seed mass $\msd$, uses a dynamical friction model for BH dynamics and executes a physical treatment of BH mergers. 
The BH population is broadly consistent with empirical constraints on the BH mass function, the bright end of the luminosity functions, and the time evolution of BH mass and accretion rate density. 
The BH mass and accretion exhibit a tight correlation with host stellar mass and star formation rate.
We trace BHs seeded before $z>10$ down to $z=3$, finding that BHs carry virtually no imprint of the initial $\msd$ except those with the smallest $\msd$, where less than 50\% of them have doubled in mass.
Gas accretion is the dominant channel for BH growth compared to BH mergers.
With dynamical friction, \asterix predicts a significant delay for BH mergers after the first encounter of a BH pair, with a typical elapse time of about $200$ Myrs. 
There are in total $4.5\times10^5$ BH mergers in \asterix at $z>3$, $\sim 10^3$ of which have X-ray detectable EM counterparts: a bright $\kpc$ scale dual AGN with $\Lx>10^{43}$ erg s$^{-1}$. 
BHs with $\mbh \sim 10^{7-8} \msun$ experience the most frequent mergers.
Galaxies that host BH mergers are unbiased tracers of the overall $\mbh-\mstar$ relation.
Massive ($>10^{11}\msun$) galaxies have a high occupation number ($\gtrsim 10$) of BHs, and hence host the majority of BH mergers.



\end{abstract}


\section{Introduction}

A plethora of observational evidence has established that supermassive black holes (SMBHs) were in place very early in the history of the Universe \citep[see, e.g.][and references therein]{Fan2019} and that they play a crucial role in many aspects of cosmic evolution. 
Observed correlations between SMBHs and various large-scale properties of their host galaxies such as galaxy stellar or bulge mass \citep[e.g.][]{Haring2004,Kormendy2013,Reines2016}, velocity dispersion \citep[e.g.][]{Ferrarese2000}, star formation rate \citep[e.g.][]{Chen2013}, etc, suggest that formation and evolution of galaxies and their central SMBHs are tightly connected.
Active galactic nuclei (AGN) powered by accreting SMBHs are believed to regulate the growth of their host galaxy and also the accretion onto the SMBH itself, leading to the co-evolution of the galaxies and their central SMBHs \citep[e.g.][]{Heckman2014}.
Understanding the formation and rapid early growth of SMBHs, as well as their interactions with their host galaxies, remains one of the most intriguing and unsolved problems in astrophysics.

The launch of next-generation telescopes in the next decade and a half will greatly advance our understanding of galaxies and SMBHs in the early Universe. 
In particular, the first galaxies will be probed and characterized by deep optical and infrared surveys using the James Webb Space Telescope~\citep[JWST,][]{JWST} and the subsequent Nancy Grace Roman Space telescope~\citep[NGRST, formerly WFIRST ][]{WFIRST}.
For the high redshift SMBHs, the large survey area of EUCLID~\citep{Euclid2019} will detect samples of the most massive and luminous BHs at near-infrared wavelengths, and the high sensitivity of JWST will probe much fainter high-redshift AGN populations in the near-infrared band.
Moreover, the proposed X-ray mission, the Lynx X-ray Observatory~\citep{Lynx2018} will provide unprecedented sensitivity to probe accreting BHs with masses $\sim 10^4 \msun$ at $z=10$, unveiling the first massive black holes in the first generation of galaxies.
Those observations will together open an electromagnetic window into the dawn of BHs and their host galaxies.

On the other hand, the first gravitational wave (GW) detections by LIGO \citep{Abbott2016} open up exciting new possibilities for the multi-messenger study of BHs.
The terrestrial interferometers of LIGO can only cover the high-frequency regime ($10-10^3$ Hz) of GW emission, and thus sources are limited to stellar-mass BH binaries with $\mbh \lsim 10^2 \msun$ \citep{Martynov2016}. Next-generation space-based missions such as the Laser Interferometer Space Antenna (LISA) \citep{LISA2017,LISA2019}, with much longer interferometer arms, will also probe lower frequency GWs ($10^{-4}-10^{0}$ Hz), and thus study coalescences of more massive BHs with $\mbh \sim 10^4-10^7 \msun$ at a wide range of redshifts with $z < 20$.
Complementary to the space and ground-based detectors, Pulsar timing array (PTAs) experiments \citep[e.g., IPTA,][]{Perera2019} making use of precisely-timed pulsars as galaxy-scale detectors would further probe GWs at nanohertz frequencies, and may be able to detect the inspiral of binary SMBHs up to $\mbh > 10^8 \msun$.
The combination of Lynx X-ray measurements of high redshift BH accretion, with GW detections of the first BH mergers, will together probe the growth of the first BHs by both accretion and mergers, unveiling a complete picture of their early assembly. 

Cosmological hydrodynamic simulations are an increasingly valuable tool for testing and modelling the formation of galaxies and SMBHs and making predictions for those upcoming observations.
Modern cosmological-scale hydrodynamic simulation projects, such as Magneticum~\citep{Hirschmann2014}, Illustris~\citep{Vogelsberger:2014}, Eagle~\citep{Schaye2015}, Horizon-AGN~\citep{Dubois2015,Volonteri2016Horizon-AGN}, MassiveBlack~\citep{Khandai2015}, BlueTides~\citep{Feng2016}, Romulus~\citep{Tremmel2017}, Illustris-TNG~\citep{Springel2018}, Simba~\citep{Dave2019-simba}, and THESAN \citep{2021arXiv211000584K, 2021arXiv211001628G,2021arXiv211002966S} have implemented ever improving sub-grid models.
They concurrently model the formation and evolution of galaxies and SMBHs through cosmic time, spanning a vast dynamic range from the creation of the cosmic web on Gpc scales to the evolution of the interstellar medium, stars and BHs on parsec scales.

In this work, we introduce results from the \asterix simulation, a large volume cosmological hydrodynamical simulation, with box size $250 \hmpc$ containing $2\times 5500^3$ particles, run to $z=3$ using the cosmological simulation code \texttt{MP-Gadget} \citep{MPGadget2018} on the Frontera supercomputer.
An earlier version of \texttt{MP-Gadget} carried out our previous large volume simulation \bluetides~that focused on studying the $z>6$ quasars and galaxies in a $400 \hmpc$ box, and has been validated against a number of observables \citep[e.g.][]{Feng2016,Wilkins2017,DiMatteo2017,Tenneti2018,Huang2018,Ni2018,Bhowmick2018,Ni2020,Marshall2020,Marshall2021}. 
Compared with the \bluetides~simulation, we have included several new physical processes in the simulation code, including models for inhomogeneous (patchy) hydrogen reionization, helium reionization, metal return from massive stars, the initial velocity offset between baryons and dark matter, as well as a dynamical friction model for BH dynamics.

Our companion paper \cite{Bird2021} discusses the predictions of \asterix for galaxy formation and shows the effects of the inhomogeneous hydrogen reionization model on star formation rates.
In this paper, we introduce the results of studies of the BH and AGN population and BH-galaxy relation, as well as the BH merger predictions made by \asterix.
Notably, with explicit inclusion of dynamical friction from collisionless particles, \asterix models massive BH mergers with a higher fidelity compared to many cosmological simulations, allowing us to make a statistical prediction for the multi-messenger signals of BH mergers.

The outline of the paper goes as follows. We give a brief introduction to the \asterix simulation in Section~\ref{section:asterix-simulation} and refer to our companion paper \cite{Bird2021} for detailed presentations of physical models for re-ionization and star formation. 
In Section~\ref{section:BH-models} we give an overview of the BH model applied in \asterix. 
In Section~\ref{section3:BH-statistics} we present statistics of the BH population and compare the simulation with several observational constraints. 
In Section~\ref{section4:BH-Galaxy} we show the connection between BHs and galaxies and also discuss the BH occupation in the galaxies. 
In Section~\ref{section5:BH-merge} we analyse the BH mergers in \asterix. 
Finally, in Section~\ref{section6:Conclusion} we summarize our results and conclude.

\section{\asterix simulation}
\label{section:asterix-simulation}

\asterix is a cosmological hydrodynamical simulation performed using a new version of the MP-Gadget simulation code, a descendant of Gadget-3. 
It contains $5500^3$ cold dark matter (DM) particles in a $250 \hmpc$ side box, and an initially equal number of SPH hydrodynamic mass elements. 
The initial conditions are set at $z=99$ and the current final redshift is $z=3$. 
The cosmological parameters used are from \citep{Planck}, with $\Omega_0=0.3089$, $\Omega_\Lambda=0.6911$, $\Omega_{\rm b}=0.0486$, $\sigma_8=0.82$, $h=0.6774$, $A_s = 2.142 \times 10^{-9}$, $n_s=0.9667$. 
The mass resolution of \asterix is $M_{\rm DM} = 6.74 \times 10^6 \hmsun$ and $M_{\rm gas} = 1.27 \times 10^6 \hmsun$ in the initial conditions. 
The gravitational softening length is $\epsilon_{\rm g} = 1.5 \hkpc$ for both DM and gas particles.

\texttt{MP-Gadget} \citep{MPGadget2018} is a massively scalable version of the cosmological structure formation code Gadget-3 \citep{Springel:2005}. 
Although many of the basic algorithms are unchanged since \cite{Feng2016}, the implementation has evolved significantly in the direction of improved speed and scalability.
Here we briefly list some of the basic features with regard to the hydrodynamic implementations and models of galaxy formation, and refer the readers to our companion paper \cite{Bird2021} for a more extensive description of the simulation code.
The sub-grid models for BHs are described in Section~\ref{section:BH-models}.

In \asterix, the gravitational dynamics uses a TreePM algorithm. 
The hydrodynamics, star formation and stellar feedback models largely follow \cite{Feng2016}.
We adopt the pressure-entropy formulation of smoothed particle hydrodynamics (pSPH) to solve the Euler equations \citep{2013MNRAS.428.2840H,2010MNRAS.405.1513R}. 
Gas is allowed to cool both radiatively following \cite{1996ApJS..105...19K} and via metal line cooling. We approximate the metal cooling rate by scaling a solar metallicity template according to the metallicity of gas particles, following \cite{Vogelsberger:2014}. 

We model patchy reionization with a spatially varying ultra-violet background using a semi-analytic method based on hydrodynamic simulations performed with radiative transfer \citep[for more details see][]{2013ApJ...776...81B}. 
For the ionized regions, we applied the ionizing ultra-violet background provided by \cite{FG2020}. 
Self-shielding is implemented following \cite{Rahmati:2013}, and ensures that gas is shielded from the ultra-violet background and thus is neutral at a density above $0.01$ atoms/cm$^{-3}$.

Star formation is implemented based on the multi-phase star formation model in \cite{2003MNRAS.339..289S}.
We model the formation of molecular hydrogen, and its effect on star formation at low metallicities, following the prescription by \cite{2011ApJ...729...36K}. 
Stars are formed with $1/4$ of the mass of a gas particle (i.e., $M_{\rm star} \sim 3\times 10^5 \hmsun$ in most cases). 
Type II supernova wind feedback is included following \cite{Okamoto2010}, assuming wind speeds proportional to the local one dimensional dark matter velocity dispersion.
Wind particles are hydrodynamically decoupled for a conditioned period of time and while decoupled do not experience or contribute to pressure forces or accrete onto a BH. 

In \asterix, we also include models for helium reionization and the effect of massive neutrinos. Metal return from AGB stars, type II supernovae and type 1a supernovae is included, following \cite{Vogelsberger2013, Pillepich2018}, although the implementation is independent and we have created our own yield tables, as detailed in \cite{Bird2021}.

We include treatment of SMBHs. Our accretion and AGN feedback models are similar to those in the \bluetides~simulation \cite{Feng:2016}, and are based on earlier work by \cite{SDH2005,DSH2005}.
Compared with \bluetides, we have changed the seeding scheme of SMBHs by drawing the BH seed mass from a power-law distribution instead of using a universal seed mass.
Furthermore, we apply the dynamic friction model \citep[tested and validated in][]{Chen2021} to allow the sinking and merger of SMBHs after a galaxy merger to be followed more accurately than with the earlier repositioning based model. 
We elaborate on our BH models in Section~\ref{section:BH-models}.

\begin{figure*}
\centering
  \includegraphics[width=1.0\textwidth]{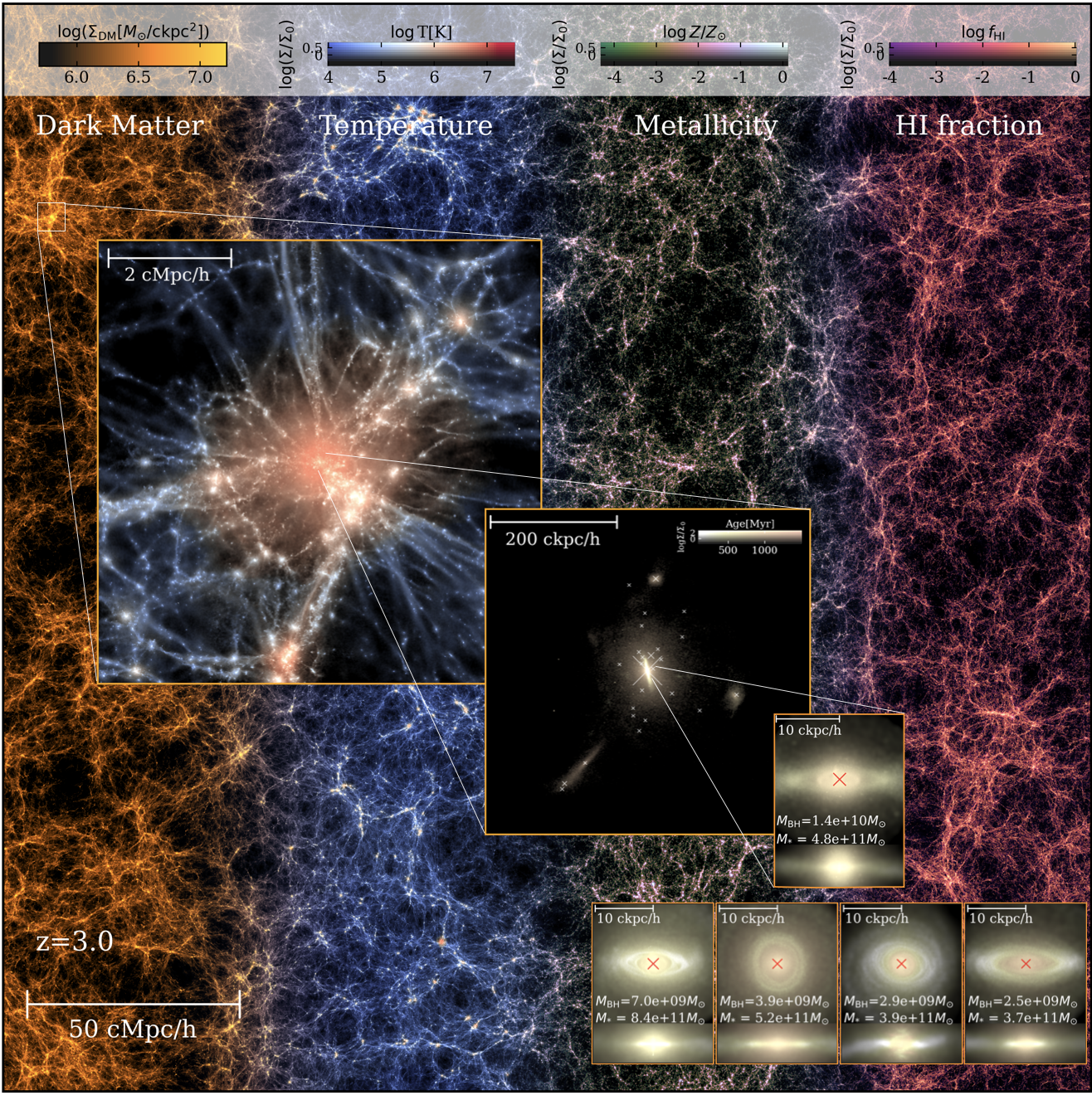}
  \caption{Illustration of the \asterix~simulation at $z=3$. 
  \textit{Background:} A $250 \hmpc \times 250 \hmpc$ field view with a slab width of $15 \hmpc$, from left to right the colour shows dark matter density (orange), gas temperature (blue), metallicity (purple) and neutral hydrogen fraction (red) respectively.
  \textit{First inset (Upper left):} Gas density field coloured by temperature, showing a $7 \hmpc$ zoomed-in region centred on a massive halo with $M_{\rm h} = 3\times10^{13} \msun$.
  \textit{Second inset:} further zoom into a $500 \hkpc$ region, showing the stellar density field centred on an ultra-massive $10^{10} \msun$ BH. The white crosses in the panel mark the positions of SMBHs in this region with the cross size scaled by the BH mass.
  \textit{Third inset:} the morphology of the host galaxy in face-on (upper panel) and edge-on (lower panel) views in a $20 \hkpc$ region around the central SMBH. Colours show the stellar age with older stars being redder.
  The bottom insets show some randomly chosen galaxies hosting $>10^9 \msun$ BHs. }
  \label{fig:Gallery}
\end{figure*}

In Figure~\ref{fig:Gallery}, we give an illustration of the \asterix simulation at $z=3$, visualizing the multi-field modelling of various physical processes for a wide range of dynamical scales.
Shown in the background is a slab of the full simulation volume with $250 \hmpc \times 250 \hmpc$ width per side and $15 \hmpc$ slab depth.
From left to right we show the projected dark matter and gas density field with colour hue set by dark matter density (orange), gas temperature (blue), metallicity (purple) and neutral hydrogen fraction (red) respectively.
For the leftmost inset we zoom into a $(7 \hmpc)^3$ region centred on a massive halo with $M_{\rm halo} = 3\times10^{13} \msun$, showing the gas density field coloured by temperature.
The second inset from the left further zooms into a $(500 \hkpc)^3$ region centred on an ultra-massive BH with $\mbh = 1.4 \times 10^{10} \msun$, residing in a host galaxy with mass $\mstar = 5 \times 10^{11} \msun$. The panel shows the stellar density field colour-coded by the age of the stars. The white crosses in the panel mark the positions of SMBHs in this region with the cross size scaled by the BH mass. 
The third inset panel shows the morphology of the host galaxy in face-on (upper panel) and edge-on (lower panel) views in a $20 \hkpc$ region around the central SMBH.
Finally, in the bottom inset, we show some randomly chosen galaxies that host $>10^9 \msun$ BHs.

\section{Black hole models}
\label{section:BH-models}

\subsection{BH seeding}

The formation mechanism of SMBH seeds is an active research topic ~\citep[see,e.g.][for recent review]{Smith2019,Volonteri2021}. Different seeding channels have been proposed, including remnants of the first generation (Population III) stars \citep{Madau2001} creating $10 \sim 10^3 \msun$ BH seeds, direct collapse of pristine (metal-free) gas \citep{Bromm2003} and runaway collapse of nuclear star clusters \citep{Davies2011} with a higher initial seed mass ($\sim 10^4 - 10^6 \msun$).
However, these processes cannot be modelled self-consistently at the resolution of current cosmological simulations. 

In \asterix, we continue to follow the practice of seeding a BH after the formation of a sufficiently massive halo. We periodically run a Friends-of-Friends (FOF) group finder during the simulation, with a linking length of $0.2$ times the mean particle separation. Haloes with a total mass and stellar mass exceeding the seeding criteria \{ $M_{\rm halo,FOF} > M_{\rm halo,thr}$; $M_{\rm *,FOF} > M_{\rm *,thr}$\} are eligible for seeding.
We apply a mass threshold value $M_{\rm halo,thr} = 5 \times 10^9 \hmsun$ and $M_{\rm *,thr} = 2 \times 10^6 \hmsun$.
The choice of $M_{\rm *,thr}$ is a conservative setting to make sure that BHs are seeded in halos with enough cold dense gas to form stars, and that there are at least some collisionless star particles in the BH neighbourhood to act as sources of dynamical friction.
The value of $M_{\rm *,thr}$ is estimated from the global $M_{\rm halo}$-$M_{\rm *}$ relation. 
Most of the FOF halos with $M_{\rm halo} > M_{\rm halo,thr}$ satisfy the $M_{\rm *,thr}$ criteria.

Considering the complex astrophysical process likely to be involved in BH seed formation, we allow haloes with the same mass to have different SMBH seeds.
Therefore, in \asterix, instead of applying a uniform seed mass for all the BHs, we probe a range of the BH seed mass $\msd$ drawn probabilistically from a power-law distribution
\begin{equation}
\label{equation:power-law}
    P(\msd ) = 
    \begin{cases}
    0 & M_{\rm sd} < M_{\rm sd,min} \\
    \mathcal{N} (M_{\rm sd})^{-n} & M_{\rm sd,min} <= M_{\rm sd} <= M_{\rm sd,max} \\
    0 & M_{\rm sd} > M_{\rm sd,max}
   \end{cases}
\end{equation}
where $\mathcal{N}$ is the normalization factor.
We set $M_{\rm sd,min} = 3 \times 10^4 \hmsun$, $M_{\rm sd,max} = 3 \times 10^5 \hmsun$, and a power-law index $n = -1$. 
In Section \ref{subsection3.4}, we investigate the effect of $M_{\rm sd}$ on BH growth. 

For each halo that satisfies the seeding criteria but does not already contain at least one SMBH particle, we convert the densest gas particle into a BH particle. 
The newborn BH particle inherits the position and velocity of its parent gas particle. 
The intrinsic mass of BH ($M_{\rm BH}$ hereafter) seeds are initially at $M_{\rm sd}$ (as a subgrid mass). However, we preserve in a separate mass variable the parent particle mass, representing a subgrid gas reservoir around the SMBH.
Neighbouring gas particles are swallowed once (Eddington-limited) accretion has allowed $M_{\rm BH}$ to grow beyond the initial parent particle mass (i.e., once it depletes the gas reservoir of the parent gas particle).

\subsection{BH dynamics and mergers}
\label{subsection:BH dynamics}

A common approach applied in cosmological simulations is to anchor the BH at the halo potential minimum by forcefully repositioning the BH particle to the local (within the SPH kernel) potential minimum at every active timestep. While this method is effective at stabilizing the BH in the halo centre, it means that the BH trajectory is discontinuous and the BH velocity is ill-defined. In addition, should two galaxies pass close enough to each other, a BH merger can result immediately even though they are not gravitationally bound.

We instead use a sub-grid dynamical friction model \citep{Chen2021} to account for the unresolved frictional force exerted by the surrounding collisionless matter (stars and stellar-mass black holes) on SMBH dynamics. In this model, SMBHs gradually sink to the galaxy centre, as their relative velocity is dissipated by dynamical friction. 
Moreover, we can make a more physical prediction for BH mergers by calculating whether the (now well-defined) relative velocities of two SMBHs allow them to be gravitationally bound.
The validation of our dynamical friction model in cosmological simulations was performed in \cite{Chen2021}. In this section, we briefly review the BH dynamics.

\subsubsection{Dynamical friction}
\label{subsubsection:dynamic-friction}

When an SMBH moves through an extended medium composed of collisionless particles with smaller mass, it experiences a drag force. The source of this force is the gravitational wake from perturbing the particles in the surrounding medium \citep{Chandrasekhar1943}. 
This drag force is called dynamical friction and it effectively transfers energy and momentum from the SMBH to surrounding particles, thus dissipating the SMBH momentum relative to its surroundings.
Dynamical friction causes the SMBH orbit to decay towards the centre of massive galaxies \citep[e.g.][]{Governato1994,Kazantzidis2005} and also stabilizes the BH motion at the centre of the galaxy.

We estimate dynamical friction on SMBHs using Eq. 8.3 of \cite{Binney2008}: 
\begin{equation}
\label{eq:df_full}
    \mathbf{F}_{\rm DF} = -16\pi^2 G^2 M_{\rm BH}^2 m_{a} \;\text{log}(\Lambda) \frac{\mathbf{v}_{\rm BH}}{v_{\rm BH}^3} \int_0^{v_{\rm BH}} dv_a v_a^2 f(v_a),
\end{equation}
$M_{\rm BH}$ is the BH mass, $\textbf{v}_{\rm BH}$ is the BH velocity relative to its surrounding medium, $m_a$ and $v_a$ are the masses and velocities of the particles surrounding the BH, and $\text{log}(\Lambda)=\text{log}(b_{\rm max}/b_{\rm min})$ is the Coulomb logarithm that accounts for the effective range of the friction between the specified $b_{\rm min}$ and $b_{\rm max}$. 
$f(v_a)$ in Eq.~\ref{eq:df_full} is the velocity distribution of the surrounding collisionless particles including both stars and dark matter. 
Here we have assumed an isotropic velocity distribution of the particles surrounding the BH so that we are left with a 1D integration. 

In \asterix, the BH seed mass extends down to $3\times 10^4 \hmsun$, which is one order of magnitude smaller than the stellar particle mass. 
At this $\mbh$ regime, the dynamical friction of BH is underestimated due to the limited mass resolution of the star particles, and so the dynamics of the seed BH would be unstable without repositioning.
As a compromise, we replace the $\mbh$ in Eq.~\ref{eq:df_full} with $M_{\rm dyn}$. This temporarily boosts the BH dynamic mass for BHs near the seed mass, stabilizing their motion during the early post-seeding evolution.
We will further discuss this implementation in Section~ \ref{subsubsection:DynamicBHMass}.

We approximate $f(v_a)$ by the Maxwellian distribution \citep[as, e.g.][]{Binney2008}, and account for the neighbouring collisionless particles up to the range of the SPH kernel of the BH particle \citep[see,][for more details]{Chen2021}. 
Eq.~\ref{eq:df_full} reduces to
\begin{equation}
    \label{eq:H14}
    \mathbf{F}_{\rm DF} = -4\pi \rho_{\rm sph} \left(\frac{GM_{\rm dyn}}{v_{\rm BH}}\right)^2  \;\text{log}(\Lambda) \mathcal{F}\left(\frac{v_{\rm BH}}{\sigma_v}\right) \frac{\bf{v}_{\rm BH}}{v_{\rm BH}}.
\end{equation}
Here $\rho_{\rm sph}$ is the density of dark matter and star particles within the SPH kernel. 
The function $\mathcal{F}$, defined as
\begin{equation}
    \label{eq:fx}
    \mathcal{F}(x) =  \text{erf}(x)-\frac{2x}{\sqrt{\pi}} e^{-x^2}, \;
    x=\frac{v_{\rm BH}}{\sigma_v}
\end{equation}
is the result of analytically integrating the Maxwellian distribution. $\sigma_v$ is the velocity dispersion of the surrounding particles.
The Coulomb logarithm $\Lambda$ is calculated with
\begin{equation}
    \Lambda = \frac{b_{\rm max}}{(GM_{\rm dyn})/v_{\rm BH}^2}
\end{equation}
where $b_{\rm max} = 20\text{ kpc}$ is an ad hoc choice for the maximum physical range to account for the frictional force. 

\subsubsection{Gas Drag}
\label{subsubsection:gas-drag}
In addition to the dynamical friction from the collisionless particles (dark matter and stars), we also add a drag force on the BH from the surrounding gas particle, with
\begin{equation}
\label{eq:drag}
    \mathbf{a}_{\rm drag} = (\mathbf{v}_{\rm gas}-\mathbf{v}_{\rm BH})\dot{M}_{\rm BH}/M_{\rm BH}
\end{equation}
This term is motivated by the BH gaining momentum from the surrounding gas continuously, bringing the sinking of the BH with respect to the gas.
It is a subdominant factor compared to the dynamical friction exerted by collisionless particles \citep{Chen2021}.

\subsubsection{Dynamic mass of BH}
\label{subsubsection:DynamicBHMass}

In \asterix, the minimal BH seed mass is $3\times 10^4 \hmsun$, orders of magnitude smaller than the stellar and DM particle masses.
Such a small BH mass relative to the surrounding particles causes noisy gravitational forces (dynamical heating) around the BH and thus instability in the BH motion.  Moreover, as shown in some previous works~\citep[e.g.][]{Tremmel2015,Pfister2019}, it is challenging to effectively model dynamical friction in a sub-grid fashion when $M_{\rm BH}/M_{\rm DM} \ll 1$.

Following \cite{Chen2021}, we introduce another BH mass tracer, the dynamical mass $M_{\rm dyn}$, to account for the force calculation of BH (including the gravitational force and dynamical friction).  
Note that we still use the intrinsic BH mass $M_{\rm BH}$ to account for the BH accretion and AGN feedback. 
When a new BH is seeded, we initialize the corresponding $M_{\rm dyn} = M_{\rm dyn,seed} = 10^7 \hmsun$, which is about $1.5 M_{\rm DM}$. 
\cite{Chen2021} showed that this alleviates dynamic heating and stabilizes the BH motion in the early growth phase.
$M_{\rm dyn}$ is kept at its seeding value until $M_{\rm BH}>M_{\rm dyn,seed}$.
After that, $M_{\rm dyn}$ grows following the BH mass accretion.

The boost of the initial $M_{\rm dyn}$ may overestimate the dynamical friction for small BHs and the resultant sinking time scale will be shortened by a factor of $\sim M_{\rm BH}/M_{\rm dyn}$ compared to the no-boost case.
On the other hand, it is also possible that the BH sinking time scale estimated in our simulation in the no-boost case could overestimate the true sinking time, as the high-density stellar bulges are not fully resolved \citep[e.g.][]{Antonini2012,Dosopoulou2017,Biernacki2017}.
Therefore, boosting the initial $M_{\rm dyn}$ seems a reasonable compromise to model the dynamics of small mass BHs while also alleviating the noisy perturbation of dynamic heating brought by the limit of resolution. 
Note that even if our dynamic friction implementation overestimates the force, it still provides a substantially more conservative estimation of BH sinking than the common model where SMBHs are aggressively repositioned to the potential minimum.

\subsubsection{BH Mergers}
\label{subsubsection:BH-merger}

In \asterix, BHs have a well-defined dissipative velocity subject to dynamic friction, allowing us to impose a more physical criterion for BH mergers based on the relative velocities and accelerations of merging SMBH pairs.
Following \cite{Bellovary2011} and \cite{Tremmel2017}, we check whether two nearby BHs are gravitationally bound using the criteria
\begin{equation}
\label{equation:merger}
    \begin{cases}
    |\bf{\Delta r}| < 2 \epsilon_{\rm g} \\
    \frac{1}{2}|\bf{\Delta v}|^2 < \bf{\Delta a} \bf{\Delta r} \,.\\
   \end{cases}
\end{equation}
Here $\epsilon_{\rm g} = 1.5 \hkpc$ is the gravitational softening length.
We set the merging distance based on $\epsilon_{\rm g}$ as BH dynamics below this distance are not spatially resolved.
$\bf{\Delta a}$,$\bf{\Delta v}$ and $\bf{\Delta r}$ denote the relative acceleration, velocity and position of the BH pair, respectively. 
Note that this expression is not strictly the total energy of the BH pair, but an approximation to the kinetic energy and the work needed to get the BH to merge. 

The BH merging criteria applied in \asterix is an improvement compared with many cosmological simulations where BH dynamics are modelled with repositioning.
In simulations with BH repositioning, the distance between the two BHs is essentially the only merging criteria. It might spuriously merge two BHs with high relative velocities when in reality they are not gravitationally bound and should not merge yet (or may never merge). 
Some cosmological simulations \citep[e.g.][]{Schaye2015,Dave2019-simba} also apply gravitational binding or escape velocity criteria for a BH merger. However, in those cases, the BHs still do not have a well-defined velocity and sinking time.

Moreover, as the BH pairs in \asterix now have well-defined orbits down to the numerical merger time, we can make use of the binary separation and eccentricity measured at the time of numerical merger as the initial conditions when post-processing BH mergers at sub-resolution scale, without having to assume a constant value as in \cite{Kelley2017}.
We will address this further in our follow up work \citep{Chen2021b}.

\subsection{BH accretion and feedback}

We model BH growth and AGN feedback as in the \bluetides~simulation. 
BHs are allowed to grow by accreting mass from nearby gas particles.
The gas accretion rate onto the BH is estimated via the Bondi-Hoyle-Lyttleton-like prescription applied to the smoothed properties of the $112$ gas particles within the SPH kernel of the BH:
\begin{equation}
\label{equation:Bondi}
    \dot{M}_{\rm B} = \frac{4 \pi \alpha G^2 M_{\rm BH}^2 \rho}{(c^2_s+v_{\rm rel}^2)^{3/2}}
\end{equation}
$c_s$ and $\rho$ are the local sound speed and density of gas, $v_{\rm rel}$ is the relative velocity of the BH with respect to the nearby gas and $\alpha = 100$ is a dimensionless fudge parameter to account for the underestimation of the accretion rate due to the unresolved cold and hot phase of the subgrid interstellar medium nearby.
Note that hydrodynamically decoupled wind particles are not included in the density calculation of Eq.~\ref{equation:Bondi}.

We allow for short periods of super-Eddington accretion in the simulation, but limit the accretion rate to $2$ times the Eddington accretion rate
\begin{equation}
\label{equation:Meddington}
    \dot{M}_{\rm Edd} = \frac{4 \pi G M_{\rm BH} m_p}{\eta \sigma_{T} c}\,.
\end{equation}
$m_p$ is the proton mass, $\sigma_T$ the Thompson cross section, c the speed of light, and $\eta=0.1$ the radiative efficiency of the accretion flow onto the BH.

In summary, the total BH accretion rate in the \asterix simulation is:
\begin{equation}
    \dot{M}_{\rm BH} = {\rm Min} (\dot{M}_{\rm B}, 2\dot{M}_{\rm Edd})
\end{equation}

Numerically, the physical BH mass, $\mbh$, grows continuously with time, while we separately track the BH dynamic mass by stochastic accretion of nearby gas particles with a probability that statistically satisfies mass conservation. Given that $M_{\rm dyn}$ is temporarily boosted (and initially fixed) for the new-born BH particles for gravity and dynamical friction calculation, we use a separate mass tracer $M_{\rm trace}$ to perform gas accretion when $\mbh < M_{\rm dyn,seed}$. This term is initialized as the parent particle mass that spawns the BH, and traces $\mbh$ with stochastic gas accretion until $\mbh >= M_{\rm dyn,seed}$. After that, $M_{\rm dyn}$ grows following $\mbh$ by swallowing gas.

To ensure stable accretion, the timestep of the BH particle is set not by the acceleration dynamics, but by being $2$ times longer than the smallest timestep found amongst the gas particles within its density kernel. 

AGN thermal feedback is implemented as follows.
The SMBH is assumed to radiate with a bolometric luminosity $L_{\rm Bol}$ proportional to the accretion rate $\dot{M}_{\rm BH}$:
\begin{equation}
\label{equation:Lbol}
    L_{\rm Bol} = \eta \dot{M}_{\rm BH} c^2
\end{equation}
with $\eta = 0.1$ being the mass-to-light conversion efficiency in an accretion disk according to \cite{Shakura1973}.
5\% of the radiated energy is thermally coupled to the surrounding gas that resides within twice the radius of the SPH smoothing kernel of the BH particle. 

The AGN feedback energy is isotropically imparted to the nearby gas particles, distributing the energy among them according to the SPH kernel.
Gas particles that are heated by AGN feedback dissipate the energy in different channels depending on their thermal properties.
Non star-forming gas cools as normal, while star-forming gas heated by the SMBH relax to the temperature corresponding to the effective equation of state for star-forming \citep[c.f.][]{2003MNRAS.339..289S}, on a time scale determined by the cooling time. Note that star-forming gas heated by other means cools to the effective equation of state on the longer relaxation timescale.

\subsection{BH catalogs in \asterix}

For every time step in the simulation, we log the full physical state of all the active BH particles, including the thermal states (density, entropy), dynamics (position, velocity, acceleration from gravity, dynamical friction, gas drag), growth state (mass, accretion rate, accretion of gas and BHs), neighbour information (number of surrounding particles, local velocity dispersion) etc, to a separate catalogue.
This high time resolution BH catalogue allows us to follow the detailed growth history of all BHs in the simulation, build the merger trees of BHs, trace the trajectory of BH merger events down to $\kpc$ scale and allow post-processing of BH mergers at sub-resolution scale based on the orbits of BHs.

\section{The global BH population}
\label{section3:BH-statistics}

\begin{figure*}
\centering
  \includegraphics[width=1.0\textwidth]{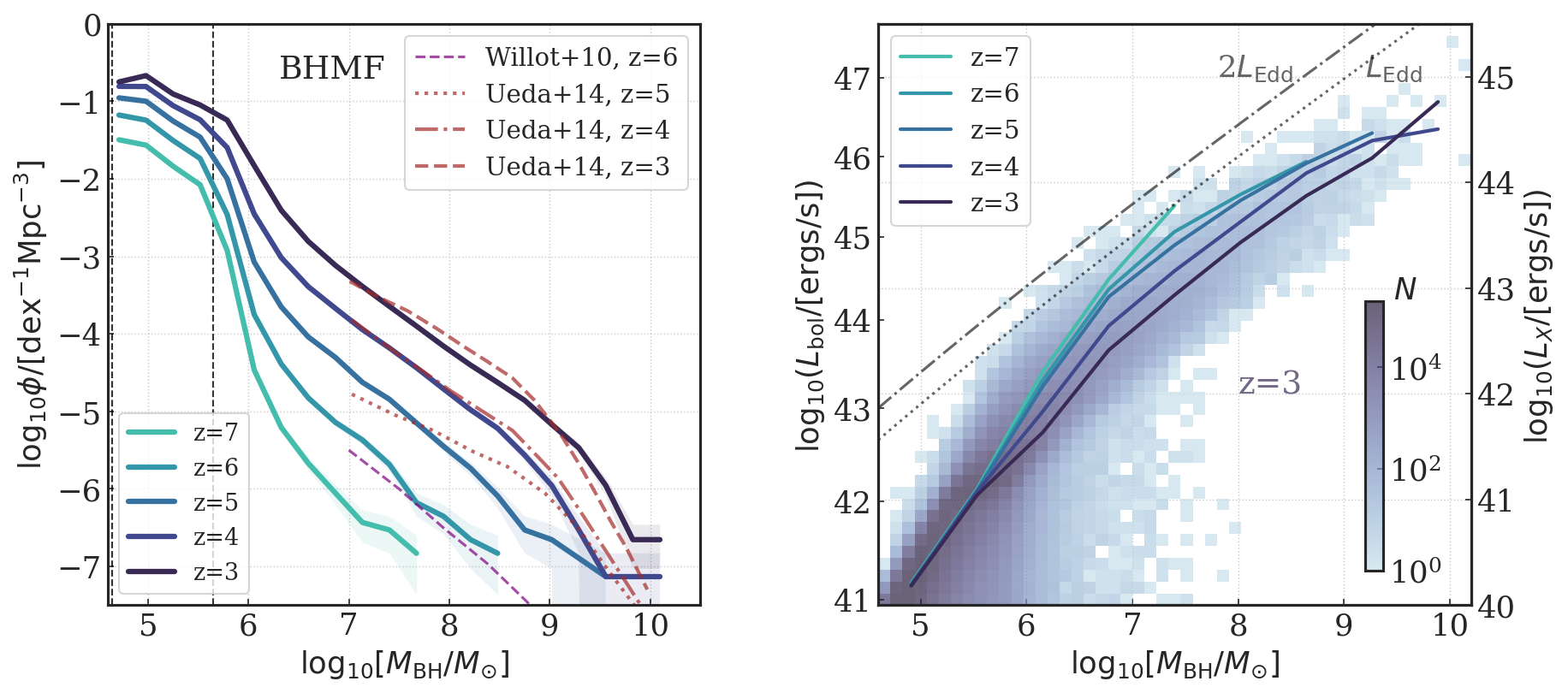}
  \caption{
  \textit{Left panel:} BH mass function in \asterix from $z=7$ to $z=3$ with coloured solid lines representing different redshifts. The grey dashed vertical lines mark the range of the BH seed mass with $M_{\rm sd,min} = 3 \times 10^4 \hmsun$, $M_{\rm sd,max} = 3 \times 10^5 \hmsun$. The purple dashed line shows an inferred BHMF at $z=6$ from \citet{Willot2010}. The brown lines give the predictions at $z=5$, $z=4$, $z=3$ by \citet{Ueda2014}, inferred from X-ray observations.
  \textit{Right panel:} Relationship between BH mass and luminosity. 
  $y$ axis gives the AGN luminosity, with $L_{\rm bol}$ given on the left axis and the corresponding hard X-ray band $\Lx$ calculated using a bolometric correction given in the right $y$ axis.
  The background shows a 2D histogram representing the $\mbh - L_{\rm bol}$ relation at $z=3$. The solid lines give the averaged AGN luminosity for each $\mbh$ bin from $z=7$ to $z=3$, with the same colour scheme as the left panel. The grey dotted-dashed and dashed lines mark $2\times L_{\rm Edd}$ and $L_{\rm Edd}$ respectively.}
  \label{fig:BHMF}
\end{figure*}


In this section, we investigate the global statistics of the BH population from $z>6$ to $z=3$ in \asterix.
\S~\ref{subsection3.1} - \S~\ref{subsection3.3} we present the statistics of the BH mass and luminosity. In
\S~\ref{subsection3.4} we consider  the global time evolution of the BH mass and accretion rate density. 
In \S~\ref{subsection:BH-seed}, we discuss the effect of the BH seeding mass on BH growth.
Finally, in \S~\ref{subsection:BH-inventory} we investigate the contributions to the BH growth from seeding, accretion and BH mergers.

\subsection{BH Mass function}
\label{subsection3.1}

We start our analysis by showing the evolution of BH mass functions (BHMFs).
In the left panel of Figure~\ref{fig:BHMF}, the coloured solid lines show the BH mass function evolving from $z=7$ to $z=3$, illustrating how the BH mass gradually builds up with cosmic time.

In \asterix, the first $10^8 \msun$, $10^9 \msun$ and $10^{10} \msun$ BHs are in place at $z = 6.6$, $z = 5.3$ and $z = 4.5$ respectively. 
With the 250 $\hmpc$ volume size of \asterix, we marginally being able to simulate the population of the first $\mbh > 10^9 \msun$ QSOs at $z>6$, which are thought to power high-$z$ quasars.
At our final redshift of $z=3$, the simulation contains a hundred $10^9 \msun$ BHs and a handful ($\sim 3$) of ultramassive BHs with $\mbh>10^{10} \msun$. 

As the first sanity check, we show in the left panel of Figure~\ref{fig:BHMF} some observational inferences of the BHMF derived from high-$z$ AGN samples.
The purple dashed line gives the estimate of the $z \sim 6$ BHMF derived in \cite{Willot2010} based on optical QSO samples at $z=6$.
The brown dotted, dashed-dotted and dashed lines are the BHMF predictions at $z=5$, $z=4$ and $z=3$ from \cite{Ueda2014},  based on X-ray observations of AGN samples (assuming mass-dependent radiative efficiency and averaged Eddington ratio $\bar{\lambda}_{\rm Edd} \sim 0.25$).
At  masses $\mbh > 10^7 \msun$, \asterix produces a $\mbh$ distribution in good agreement with these observational predictions from $z=6$ to $z=3$.
A caveat is that there are large uncertainties (and inconsistencies with the simulation model) lying in the assumptions applied by these observational inferences of BHMFs. 
The match to within an order of magnitude across a wide range of $\mbh$ and redshift still validates that the BH model in \asterix seems to be building up the $\mbh$ population in a reasonable way.


BH populations at the small mass end are sensitive to the seeding prescription and to the level of growth achieved before reaching AGN feedback modulated self-regulation.
In \asterix, the BH population with  $\mbh < 10^6 \msun$ is dominated by the seed BHs. 
The dashed vertical lines mark the range of the BH seed with $M_{\rm sd,min} = 3 \times 10^4 \hmsun$ ($4.4 \times 10^4 \msun$), $M_{\rm sd,max} = 3 \times 10^5 \hmsun$ ($4.4 \times 10^5 \msun$). 
Instead of spiking at a uniform $\msd$, the power-law seeding prescription applied in \asterix allows the BH population to extend down to a lower $\msd$ with a flatter slope, and allows a statistical diagnosis of the impact of $\msd$  on  BH growth, which will be further discussed in Section~\ref{subsection:BH-seed}. 
We will show there that $\msd$ has a limited impact on the higher mass BH population, for which growth is mostly dominated by the local environment, governing gas accretion.


\subsection{Relation between BH mass and luminosity}
\label{subsection3.2}

Now we investigate the relationship between $\mbh$ and luminosity across different redshifts. 
The AGN luminosity is calculated from the BH accretion rate using bolometric luminosity corrections.
We calculate the bolometric luminosity of AGN via Eq.~\ref{equation:Lbol} with $L_{\rm Bol} \propto \dot{M}_{\rm BH}$ and the assumption of a uniform mass-to-light conversion efficiency of $\eta = 0.1$.
To compare with AGN observations in different bands such as the X-ray and UV, we apply the corresponding bolometric corrections.

In particular, the X-ray band luminosity is estimated by converting AGN $L_{\rm Bol}$ to the luminosity in the hard X-ray band [2-10]~keV following the bolometric correction $L_X = L_{\rm Bol}/k$ from \cite{Hopkins2007}
with
\begin{equation}
\label{equation:LbolLx}
    k(L_{\rm Bol}) = 10.83(\frac{L_{\rm Bol}}{10^{10}L_\odot})^{0.28}+6.08(\frac{L_{\rm Bol}}{10^{10}L_\odot})^{-0.020}\,.
\end{equation}
Here $L_{\rm Bol}$ is given by Eq.~\ref{equation:Lbol}, and $L_{\odot}$ refers to the bolometric solar luminosity $L_{\odot}=3.9 \times 10^{33}$~erg s$^{-1}$.

To calculate the UV band AGN luminosity, we convert $L_{\rm Bol}$ to rest frame UV band absolute magnitude $M_{\rm UV}$ following the bolometric corrections from \cite{Fontanot2012}
\begin{equation}
\label{equation:LbolMuv}
    M_{\rm UV,AGN} = -2.5 \rm{log_{10}} \frac{L_{\rm Bol}}{f_B \mu_B}+34.1+\Delta_{\rm {B,UV}}\,,
\end{equation}
where $f_{B}=10.2$, $\mu_B = 6.7 \times 10^{14}$Hz and $\Delta_{\mathrm{B,UV}}$=-0.48.

The right panel of Figure~\ref{fig:BHMF} shows the distribution of $\mbh$ versus the AGN luminosity. 
The left $y$-axis is labelled in units of bolometric luminosity, while the right $y$-axis marks the corresponding $\Lx$ calculated via Eq.~\ref{equation:LbolLx}.
The background shows the 2D histogram of the $L_{\rm BH} - \mbh$ relation for the BH population at $z=3$, with the purple solid line giving the mean value of $L_{\rm BH}$ for each $\mbh$ bin.
We also give the average $L_{\rm BH} - \mbh$ for higher redshifts from $z=7$ to $z=3$ (using coloured solid lines with the same colour scheme as the left panel) to show time evolution.

At the low mass end, $\mbh \lsim 10^7 \msun$, $L_{\rm BH}$ exhibits a steep positive correlation with $\mbh$ as the BH growth is dominated by the initial Bondi-Hoyle-like accretion. More massive BHs experience a higher accretion rate, which also gives on average a higher Eddington accretion ratio $\lambda_{\rm Edd} \equiv \dot{M}/\dot{M}_{\rm Edd}$.
As shown by the $L_{\rm BH}-\mbh$ distribution at $z=3$ in the background, some of the BHs with $\mbh \sim [10^6, 10^7] \msun$ can temporarily reach Eddington-limited accretion (marked with a grey dashed-dotted line).
At increased mass, $\mbh \gtrsim 10^7 \msun$, the $L_{\rm BH}-\mbh$ curve flattens, making the overall Eddington accretion ratio lower for more massive BHs.
This signifies that BH growth is self-regulated by AGN feedback, with massive BHs dumping thermal energy into the surrounding gas and the resulting hot and 
sparse gas slowing down the BH accretion. 
The overall $\lambda_{\rm Edd}$ tends to peak at $\mbh \sim 10^7 \msun$, a typical feature found in Bondi-Hoyle-like BH accretion models \citep[see, e.g.][]{DeGraf2012}.

As shown by the coloured solid lines, the mean $L_{\rm BH}-\mbh$ relation exhibits a clear time evolution across different redshifts, with the mean Eddington accretion ratio for BHs in the same $\mbh$ regime decreasing when going to lower redshifts. 
For $\mbh \sim 10^8 \msun$, the average $\lambda_{\rm Edd} \sim 0.07$ at $z=3$, $\sim 0.12$ at $z=4$, and $\sim 0.2$ at $z=5\sim 6$.
The decrease in $\lambda_{\rm Edd}$ with redshift is a consequence of persistent AGN feedback and lower cold gas fractions in BH environments at lower redshift \citep[see discussions in, e.g.,][]{Angles-Alcazar2015}, and is mirrored in a similar decrease in star formation \citep[see][]{Bird2021}. 
As we will show in Section~\ref{subsubsection:obscuration}, AGN at higher redshift is typically embedded in denser gas environments.
Observational evidence also suggests an increasing $\lambda_{\rm Edd}$ at higher redshift \citep[e.g.][]{Lusso2012,Aird2015}, supporting the trend found in the simulation.

\begin{figure*}
\centering
  \includegraphics[width=1.0\textwidth]{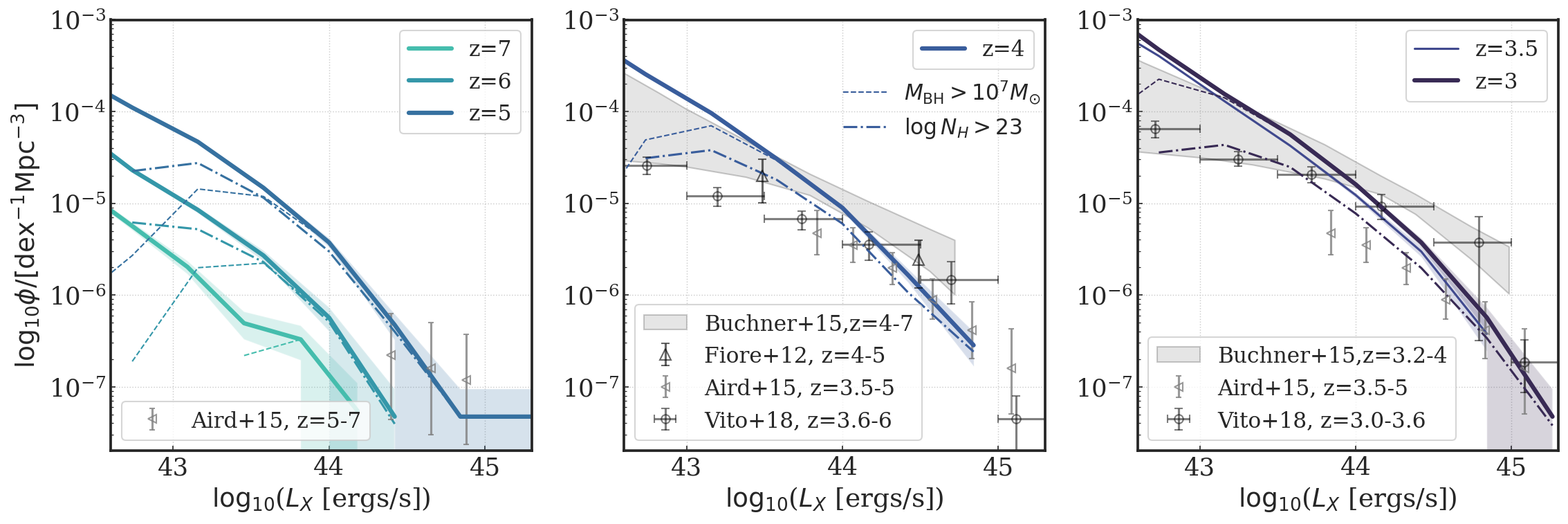}
  \caption{
  Predictions for the X-ray band luminosity function (XLF) of AGN from $z=7 \sim 5$ (first panel), $z=4$ (second panel) and $z=3$ (third panel), for comparison with observations. 
  Grey data points with error bars are the X-ray observational results from \protect\cite{Fiore2012}, \protect\cite{Aird2015}, and \protect\cite{Vito2018}. The grey shaded area covers the 90\% credible interval inferred from X-ray selected AGN samples reported by \protect\cite{Buchner2015}. The coloured dashed lines show the XLF contributed by the $\mbh > 10^7 \msun$ BH population at each redshift. The dashed-dotted lines show the contribution of the obscured AGN population embedded in local gas column density with $N_{\rm H} > 10^{23}$ $\mathrm{cm}^{-2}$ (see text for more details).}
  \label{fig:BH-XLF}
\end{figure*}

\subsection{BH Luminosity Functions}
\label{subsection3.3}

In this section, we discuss the predictions for the AGN luminosity functions (in X-ray and UV bands) for the BH population in \asterix and compare them to current observational constraints at $z\sim 3-7$.
The AGN luminosity functions (LFs: 1-point functions) provide basic statistics to gauge the accretion history of BHs as tracked in our simulation and hence are a fundamental way to validate our BH model.
In addition, following \cite{Ni2020}, we calculate the hydrogen column density, $\NH$ due to the gas in the host galaxies and assess its role in AGN obscuration. 
Although we cannot resolve scales relevant for the obscuring AGN torus, galactic obscuration serves as a lower limit on the total amount of obscuration expected. 
We compare our derived galactic $\NH$ as a function of X-ray luminosity to the current obscured fractions of AGN inferred from X-ray observations at $z > 3$.

\subsubsection{X-ray band AGN luminosity function}

In Figure~\ref{fig:BH-XLF} we show the X-ray luminosity function (XLF) from $z = 7, 6$ and $z = 5$ (left panel) and at $z =4$ and $z=3$ in the middle and right panels respectively. 
At the highest redshift ($z=5-7$), the XLF has been observationally constrained only at the very bright end ($\log \Lx \gtrsim 44$). 
At $z=4$ and $z=3$ the measurements extend down by an order of magnitude in luminosity, close to $\log \Lx \sim 43 $ (corresponding to the flux limit of the deepest Chandra fields). 
The grey data points in Figure~\ref{fig:BH-XLF} are a collection of observational constraints reported by \cite{Fiore2012}, \cite{Aird2015}, \cite{Vito2018} in different redshift ranges.
The grey shaded band covers the 90\% interval determined from AGN samples reported by \cite{Buchner2015} through a non-parametric approach used to infer the intrinsic LF and include a correction for obscured ($\NH = 10^{22-24} \mathrm{cm}^{-2}$) and Compton thick ($\NH > 10^{24} \mathrm{cm}^{-2}$) AGN.

Overall the AGN LF at $\log \Lx > 43.5$ shows good agreement compared to the observational data. 
When we consider the $\log \Lx < 43$ population in \asterix, the AGN abundance is about $0.5 \sim 1.0$ dex higher compared to observational constraints provided by \cite{Vito2018}, and is close to the upper bound of confidence interval predicted by \cite{Buchner2015}.  
We note that the \asterix XLF determined using the full BH population has a steeper slope at the faint end with $\log \Lx \lsim 43.5$. 

The details (and excess) at the faint end of the LF are a common issue seen in many cosmological simulations \citep[e.g., see discussions in][]{Sijacki2015, Volonteri2016Horizon-AGN}.
We note that there are large uncertainties related to the comparison with the observed faint end AGN population, such as (i) observations of AGN at the faint end could be strongly influenced by gas obscuration, (ii) simulation modelling of small-mass BHs is more uncertain than for larger BHs, (iii) several assumptions are applied to model the simulated AGN luminosity such as radiative efficiencies and bolometric corrections.

Dashed lines in Figure~\ref{fig:BH-XLF} show the XLF  for the subset of BHs with $\mbh > 10^{7} \msun$. 
We can see that the $\log \Lx < 43$ AGN population is dominated by smaller mass BHs with $\mbh < 10^7 \msun$ which have not yet reached the self-regulated regime and might be subject to the
 uncertainties lying in the modelling of small mass BHs.
Therefore, exclusion of small mass BHs or BHs residing in small halos can provide a better agreement at the faint end (for example,\cite{Sijacki2015} applied $\mbh > 5 \times 10^7 \msun$).

It has now been well established from X-ray studies that the vast majority of AGN are obscured by thick columns of gas and dust, and the predictions for the XLF, particularly at the faint end, are sensitive to the amount of obscuration to the AGN. 
In the next section, we focus the discussion on the important effects of obscuration and the role it plays in predictions of the LFs.

\begin{figure*}
\centering
  \includegraphics[width=1.0\textwidth]{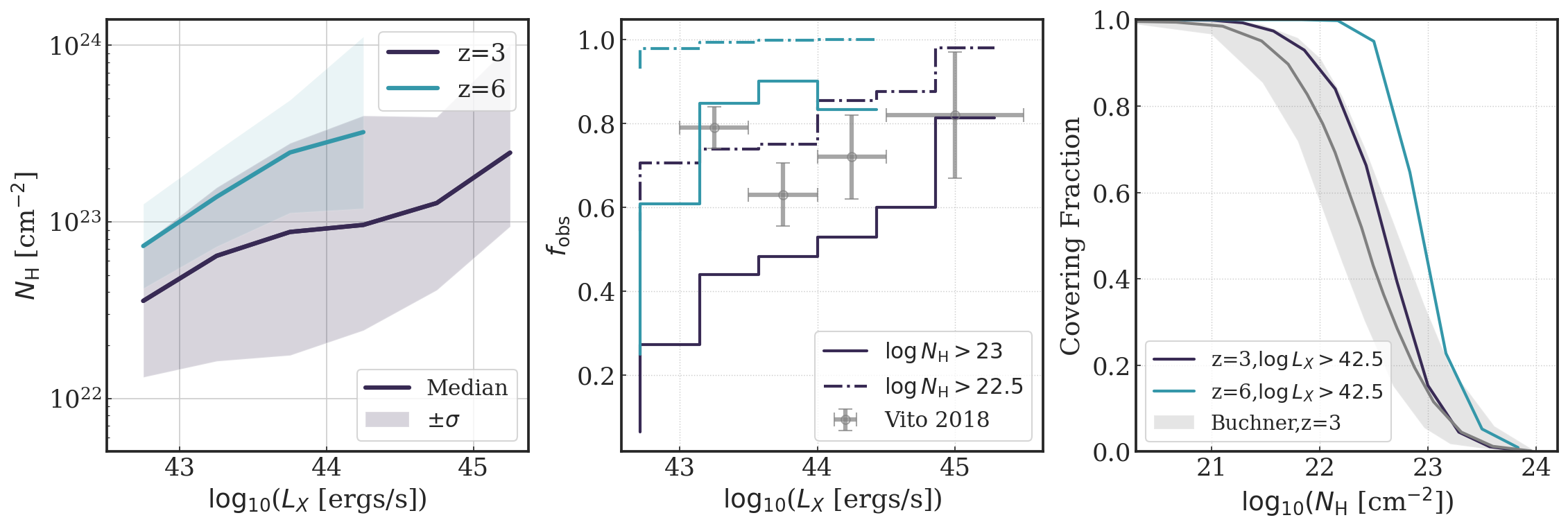}
  \caption{
  \textit{Left panel:} The relation between gas column density $\NH$ and AGN luminosity for the AGN population with $\log \Lx > 42.5$ at $z=6$ and $z=3$. For each AGN we compute $\NH$ along 48 random sight lines. The solid line gives the running median, with the shaded area covering the 16 - 84th percentile of the $\NH$ distribution in each $\Lx$ bin. 
  \textit{Middle panel:} Binned estimates of the obscured AGN fraction as a function of $\Lx$ shown at $z=6$ and $z=3$. Solid and dashed-dotted lines show the fraction of sightlines with $\NH > 10^{23}$ $\rm{cm}^{-2}$ and $\NH > 10^{22.5}$ $\rm{cm}^{-2}$ respectively. 
  The obscured fraction is calculated based on all lines of sight for the AGN populations in the corresponding $\Lx$ bin. Grey data points with error bars are the observational results of \protect\citet{Vito2018}, based on the AGN population at redshift $z=3-6$.
  \textit{Right panel:} Fraction of AGN that is covered by a given column density $\NH$, calculated based on all lines of sight for the $\log \Lx > 42.5$ AGN population at $z=6$ and $z=3$. The grey shaded area shows predictions for galaxy-scale obscuration of $z \sim 3$ AGN from \protect\citet{Buchner2017b}.}
  \label{fig:AGN-obsc}
\end{figure*}

\begin{figure}
\centering
  \includegraphics[width=1.0\columnwidth]{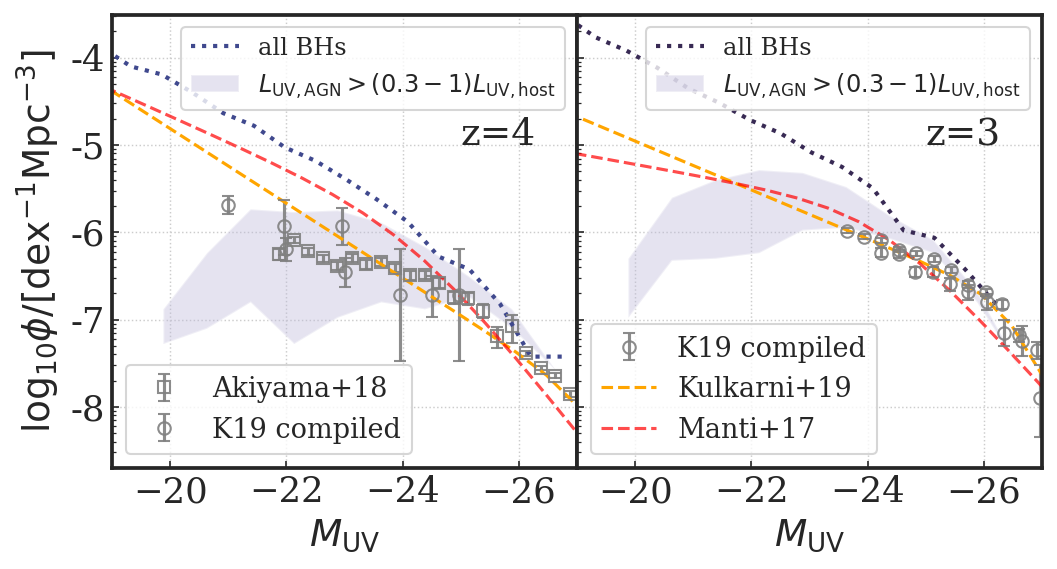}
  \caption{UV band luminosity function of AGN populations at $z=4$ and $z=3$. The blue line gives the UV luminosity function for all BH populations. The purple shaded area gives the area bounded by the UVLF from the AGN population with $L > 0.3 \times L_{\mathrm{UV, host}}$ (the host galaxy) and the AGN population with $L > 1 \times L_{\mathrm{UV, host}}$. 
  The grey data points give the observational data collected from \protect\cite{Kulkarni2019} and also \protect\cite{Akiyama2018}.
  The red and orange dashed lines give the fits to observed AGN UVLF reported by \protect\citet{Manti2017,Kulkarni2019}.}
  \label{fig:BH-UVLF}
\end{figure}

\subsubsection{The fraction of AGN obscured by galaxy-scale gas in the host}
\label{subsubsection:obscuration}

X-ray surveys over the last decade indicate that $20-40$\% of AGN are hidden behind Compton-thick column densities (with $\NH > 10^{24}$ $\mathrm{cm}^{-2}$) and, of the remaining population, $\sim 75$\% are obscured, with $\NH = 10^{22} \sim 10^{24} \mathrm{cm}^{-2}$ \citep[e.g.][]{Ueda2014,Buchner2015,Aird2015} at the peak of AGN activity at redshift $z=0.5-3$.
Up to $z \sim 3$, the highest obscured fractions are found at the faint end \citep[e.g.][]{Ueda2014,Vito2018}, showing that $80\%$ of the AGN population at $z\sim 3-7$ is obscured with little dependence on X-ray luminosity.

Traditionally, AGN obscuration is associated with the “torus”, a nuclear structure (on scales of $\sim 10$ pc) surrounding the accretion disk of the BH, which is completely unresolved in cosmological simulations. 
However, we can directly assess the contribution to the obscuration due to the gas in the host galaxy \citep[mostly due to the same, cold gas reservoir which is responsible for fuelling the AGN: see][]{Ni2020} without accounting for nuclear torus obscuration.
The obscured fraction of high-$z$ AGN is constrained to be significant \citep[see, e.g.][]{Vito2018}.
Our estimated galactic scale $\NH$ can separate the large-scale and small-scale obscured contribution for the high-$z$ AGN population. We also use it to assess the predicted contribution of galactic scale obscuration in the high redshift regime, where galactic gas densities can be high in AGN host galaxies.

We calculate gas column density $N_{\rm H}$ for each AGN from different lines of sight following \cite{Ni2020}, analysing all AGN  with $\Lx > 10^{42.5}$ erg s$^{-1}$.
For each AGN, we cast 48 evenly distributed lines of sight starting from the BH position and calculate the hydrogen column density $N_{\rm H}$ along each sightline.
More specifically, we first determine the neutral hydrogen gas density field using the SPH formalism 
\begin{equation}
\rho(\mathbf{r}_i) = X \sum_{j} \chi_{j} m_j W_{ij} =  X \sum_{j} \chi_{j} m_j W(|\mathbf{r}_i - \mathbf{r}_j|,h_j),
\label{equation:rho_i}
\end{equation}
where $\sum_j$ is a sum over all the neighbouring gas particles within the smoothing length $h$, $W$ is the quintic kernel for density estimation, $X = 0.76$ is the hydrogen mass fraction and $\chi$ is the neutral hydrogen fraction for gas particles.
We then calculate $N_{\rm H}$ by integrating the (neutral) hydrogen number density along each line of sight: 
\begin{equation}
\label{equation:NH}
    N_{\rm H} =  \int_{\rm ray} \rho(l)/m_p dl
\end{equation}
with $m_p$ the proton mass.

The dashed-dotted lines in Figure~\ref{fig:BH-XLF} show the AGN population with $\NH>10^{23} \mathrm{cm}^{-2}$. The $\NH$ threshold follows the convention of \cite{Vito2018}.
Note that $\NH$ calculated here is the galactic obscuration only and thus a lower limit on the total obscuration, as we cannot resolve the nuclear scale "torus".
The first panel of Figure~\ref{fig:AGN-obsc} shows the relation between the X-ray absorbing column densities $\NH$ due to gas in the galaxy, and AGN activity $\Lx$.
We find that galaxy-scale obscuration is important at column densities $\NH = 10^{23-24}$ cm$^{-2}$ at $z=6$.
At $z=3$, however, the typical $\NH$ decreases to $10^{22-23}$ cm$^{-2}$, as a result of the lower gas density in galaxies.

The second panel shows the obscured fraction $f_{\rm obs}$ of AGN at $z=3$ and $z=6$ as function of $\Lx$. The grey data points show constraints from \cite{Vito2018} for the fraction of AGN with $\NH>10^{23}$ cm$^{-2}$ at $z=3-6$.
At both redshifts, our estimated $\NH$ or the obscured fraction $f_{\rm obs}$ increases with $\Lx$: this is in contrast with the flat or decreasing trends shown in the observed relations at $z>3$ \citep[e.g.][]{Ueda2014,Vito2018}.
At $z=6$, galactic-scale gas can provide Compton thick column densities at large luminosities, consistent with the observationally inferred obscured fraction from \cite{Vito2018}. 
However, at $z=3$, it is not possible to reproduce the number of Compton thick AGN through host-galaxy obscuration. 
This implies that Compton-thick obscuration is associated with  nuclear obscuration in the unresolved vicinity of the BH by the time $z=3$ is reached, particularly at low luminosities.
What we classify as an 'unobscured' AGN population in \asterix is therefore likely to be greatly overestimated.

We can conclude that most Compton-thick obscuration can be associated with galactic scale gas at $z\sim 6$ (at least at high luminosities). 
However, at $z=3$, a large fraction of it needs to be associated with the nuclear regions, at least for AGNs $\log \Lx < 43.5$.
By $z=3$, the average gas density in galaxies has decreased and only massive galaxies reach gas densities which can give rise to significant absorption. 
We will further discuss this in Section~\ref{subsection: NH-mstar}, where we show the column density to the AGN as a function of the host galaxy stellar mass. 
At $z=3$, $\NH$ is below $10^{24}$ cm$^{-2}$ for most galaxies.

The third panel of Figure~\ref{fig:AGN-obsc} shows the covering fraction for $\log \Lx > 42.5$ AGN as a function of $\NH$ at $z=6$ and $z=3$.
The grey shaded area shows the range of inferred galaxy-scale obscuration allowed at $z=3$ in a model based on observations of gamma-ray bursts (GRBs) reported by \cite{Buchner2017b}.
\asterix predicts covering fractions of galactic obscuration broadly consistent with the observational inferences, indicating that the galactic level obscuration in the simulation is reasonable.
While galactic obscuration does not produce Compton-thick column densities, it can provide high covering fractions (50 - 90\% at $z=6$, decreasing to $20-60$\% at $z=3$) at $\NH = 10^{23}$ cm$^{-2}$.
The high covering fractions suggest that a substantial part of the obscured/unobscured dichotomy can still be caused by galaxy-scale gas at high-$z$ and high luminosities. 

Our main result is that the host galaxy gas provides a positive correlation between luminosity and obscuration, implying that nuclear absorption still plays the most significant role at low luminosities as $z$ decreases from $z=6$ to $z=3$. 
However, X-ray inferred column densities imply an opposite correlation, with faint AGN experiencing the largest $\NH$.
As a result, we cannot directly assess the extent to which \asterix produces an excess of AGN at the faint end of the XLF.
With insufficient $\NH$ (without the torus) in the simulation, we are unable to properly represent the typically heavily obscured / optically-thick population at the faint end of the LF.

\subsubsection{UV band AGN luminosity function}

Figure~\ref{fig:BH-UVLF} shows the predicted, rest frame, UV luminosity function for the AGN in \asterix. The UV band AGN luminosity is calculated using the bolometric correction in Eq.~\ref{equation:LbolMuv}. 
We compare the predictions with observational results from the Hyper Suprime-Cam Wide Survey \citep{Akiyama2018}, which currently provides the best constraints on the faint end of the UV LF of AGN at $z=4$. We also use the UV observational data compiled by \cite{Kulkarni2019} and show (red and orange dashed lines) two fits to AGN UVLF reported by \cite{Manti2017,Kulkarni2019}.
The \asterix bright AGN population agrees reasonably well with the observational constraints at $z=3$ and $z=4$.

The UVLF is extremely sensitive to dust extinction and the uncertainties in the obscuration corrections are large. 
These effects could easily result in an order of magnitude or more uncertainty in AGN number density in the UV band \citep[see also][]{Ni2020}. 
One approach to estimate the dust obscuration is to convert the hydrogen column density $\NH$ to dust abundance with the assumption of a (constant) gas-to-dust ratio and use a dust interaction cross-section (calibrated on, e.g. the Small Magellanic Cloud model) to account for UV band extinction. 
However, in this work, we find that these assumptions, along with our galactic gas $\NH$, under-predict the UVLF of AGN and hence lead to an inconsistency between UV and X-ray LFs. This indicates that the extinction in the optical/UV is over-predicted by a model with the
constant dust-to-gas ratio and the SMC-like extinction curve. 
This result supports a similar conclusion from the empirical modelling by \citet{Shen2020}. For this reason, that work adopts a redshift-dependent dust-to-gas ratio and a  Milky Way-like extinction curve which is shallower than the SMC-like curve.

There remains much uncertainty as to how to calculate AGN obscuration in the UV band and different empirical models have been adopted  \citep[e.g. see recent discussions in][]{Shen2020,Yung2021} that all require a somewhat suppressed effect of dust for AGN at high-$z$. 
Given that our model does not account for nuclear absorption, there are large uncertainties in the dust modelling and the faint end of the UVLFs remains largely unconstrained.
Here we discuss the effects on UVLF by considering AGN UV luminosity compared to their host galaxy.
In particular, we show the contribution to the UVLF due to AGN that are at least 30\% of the UV luminosity of their host galaxies.

We calculate the UV luminosity of the host galaxy by modelling the spectral energy distribution (SED) of a simple stellar population (SSP) to each star particle based on its age and metallicity which is discussed in detail and compared to the galaxy UVLF in our companion paper \cite{Bird2021}.

The shaded area in Figure~\ref{fig:BH-UVLF} gives the area bounded by the UVLF from the AGN population with $L_{\mathrm{UV,AGN}} > 0.3 \times L_{\mathrm{UV, host}}$ (the host galaxy) and the AGN population with $L_{\mathrm{UV,AGN}} > 1 \times L_{\mathrm{UV, host}}$.  
Interestingly, this band shows a broad consistency with observations. 
We find that at $M_{\mathrm{UV}} > -23$, most of the AGN have a UV luminosity fainter their host galaxy. 
Excluding these makes the remaining UVLF $>1$ dex lower than the overall AGN population.
Detection of AGN in this range of UV magnitudes is likely very challenging due to the dominant contribution of the host galaxies.
This exercise demonstrates another potential important effect that may tend to suppress the number of faint AGN detected by the UV observations at high redshifts. 
Similar results were discussed in the analysis by \cite{Volonteri2017} at $z=6$.
To conclude, \asterix predicts that at the bright end of the UV magnitude ($M_{\rm UV} < -24$), a significant amount of the AGN is brighter than their host galaxies. 
While at the faint end of UVLFs, the host galaxies can be a large contaminant to observations.

\begin{figure*}
\centering
  \includegraphics[width=1.0\textwidth]{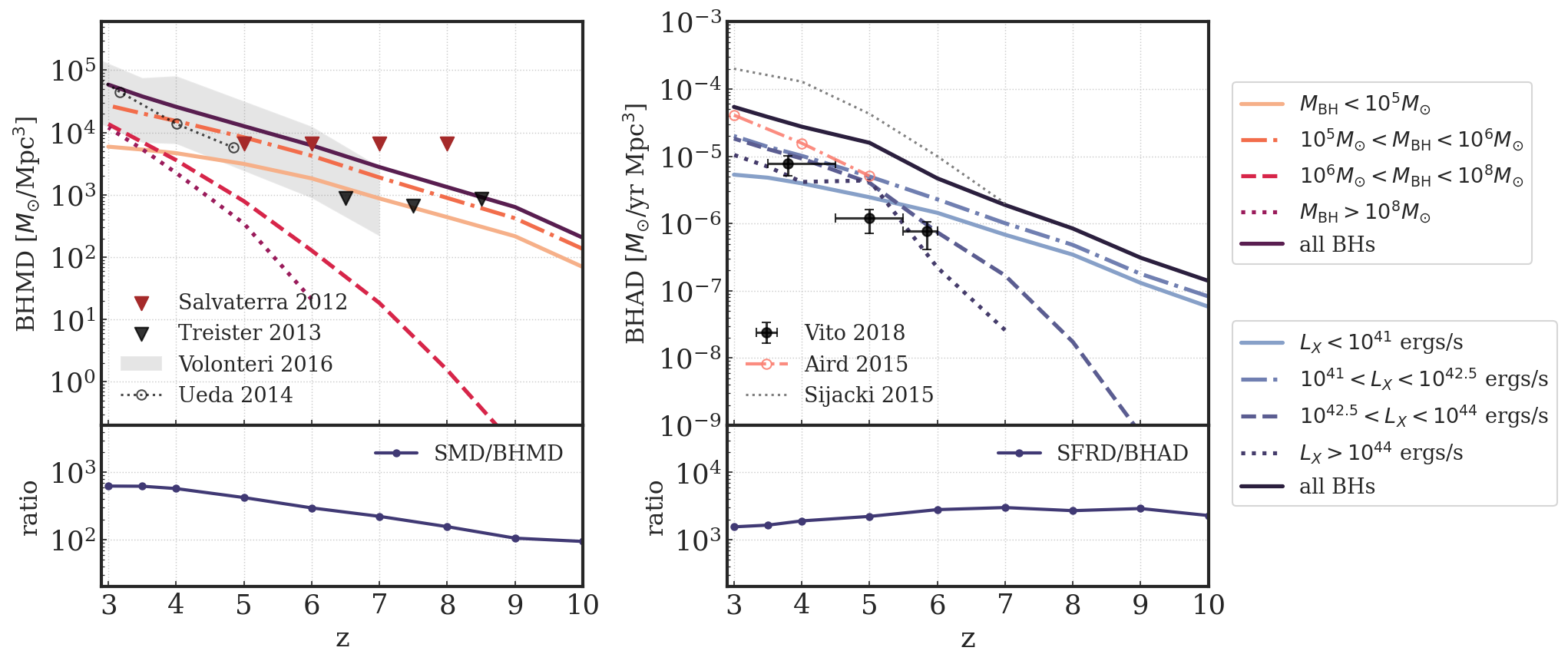}
  \caption{Evolution of BH mass and accretion rate densities.
  \textit{Left panel}: The upper panel shows the evolution of the global BH mass density (BHMD) as a function of redshift. The black solid line includes all BHs. Coloured lines show the BHMD contributed by the BHs in a given range of masses, as indicated by the legend on the right. 
  The brown and black triangles are upper limits inferred from \protect\citep{Salvaterra2012,Treister2013}. 
  We also show prediction of BHMD from \protect\citet{Ueda2014} and \protect\citet{Volonteri2016}.
  The lower panel shows the ratio between the global stellar mass density (SMD) and BHMD.
  \textit{Right panel}: The upper panel shows the evolution of the BH accretion rate density (BHAD). Coloured lines show the BHAD contributed by BHs from different ranges of X-ray luminosity as indicated by the legend on the right.  
  The lower panel shows the ratio of the global star formation rate density (SFRD) to the BHAD.
  Black and pink data points show X-ray observational prediction from \protect\citet{Vito2018} and \protect\citet{Aird2015}.
  The grey dotted line gives the simulation prediction from Illustris \protect\citep{Sijacki2015}.}
  \label{fig:BHrate-history}
\end{figure*}

\subsection{BH Mass density and accretion rate density}
\label{subsection3.4}

This section discusses the global evolution history of the BH mass density (BHMD) and BH accretion rate density (BHAD) from $z=10$ to $z=3$.
The left panel of Figure~\ref{fig:BHrate-history} shows the BHMD as a function of redshift. The red to orange coloured lines denote the BHMD contributed by BHs in different ranges of  $M_{\rm BH}$ (as indicated in the legend).
The solid triangles indicate the upper limits derived from deep X-ray observations \citep{Treister2013} or from the integrated X-ray background \citep{Salvaterra2012}.
Note that those X-ray inferred upper limits do not include Compton-thick AGNs and might thus miss the contributions from a fraction of BHs.
The black dotted line marked by empty circles was derived by \cite{Ueda2014}  from observational constraints on the AGN XLF.  
The grey shaded area denotes the predicted range of BHMD for BHs with $\mbh > 10^5 \msun$ reported by \cite{Volonteri2016}, which is derived from the galaxy mass function \citep{Grazian2015} after making assumptions about the BH-stellar mass relationships. 
We can see that the \asterix BHMD is in overall good agreement with the observational constraints.
BHs within the mass range [$10^5$, $10^6$] $\msun$ make the dominant contribution to BHMD.

The right panel of Figure~\ref{fig:BHrate-history} shows the BHAD with different coloured lines showing contributions from different $\Lx$ bins. 
At $z>5$ the BHAD is dominated by AGN with relatively low luminosity in the range of $\log \Lx$ = [41, 42.5] erg s$^{-1}$, while at lower redshift ($z<5$) more luminous AGNs with $\log \Lx > 42.5$ start to dominate the BH accretion rate.
The black data points show the observational results from \cite{Vito2018}, derived from samples of X-ray-detected AGN. 
As a comparison, we also show the simulation result from Illustris \citep{Sijacki2015} (including all BHs) as the black dotted line.
The results  of \cite{Vito2018} effectively probe AGN down to the luminosity limit $\log \Lx > 42.5$ for $z = 3 \sim 6$, but beyond this ($\log \Lx < 42.5$) the sample is strongly affected by incompleteness.
We can see that the dashed and dotted lines (corresponding to the $\log \Lx > 42.5$ AGN population) show overall good agreement with \cite{Vito2018} at $z=6$ and at $z \sim 4$, but overshoot by about 0.5 dex at $z=5$.


The lower left panel shows the ratio of the stellar mass density (SMD) to the BHMD at each redshift and the lower right panel shows star formation rate density (SFRD) divided by the BHAD.
Here SMD and SFRD are all calculated based on the galaxies with $M_* > 10^7 \msun$ in the simulation. 
Since the star formation process builds up galaxy mass faster than  BH formation and accretion (as also indicated by the SFRD/BHAD ratio), the ratio between SMD and BHMD slowly increases with time, from $\sim 100$ at $z=10$ to $\sim 700$ at $z=3$.
However, we note that the ratio of SFRD and BHAD does not show very strong time evolution, with SFRD/BHAD lying between $1000 - 3000$ for all redshifts.
Some observations \citep[e.g.][]{Mullaney2012} and simulations \citep[e.g.][]{Ricarte2019,Thomas2019-simba} also exhibit no strong redshift evolution of the relation between star formation rate and BH accretion rate, and treat this as evidence for or a consequence of BH-galaxy coevolution.


\begin{figure*}
  \centering
  \includegraphics[width=0.95\textwidth]{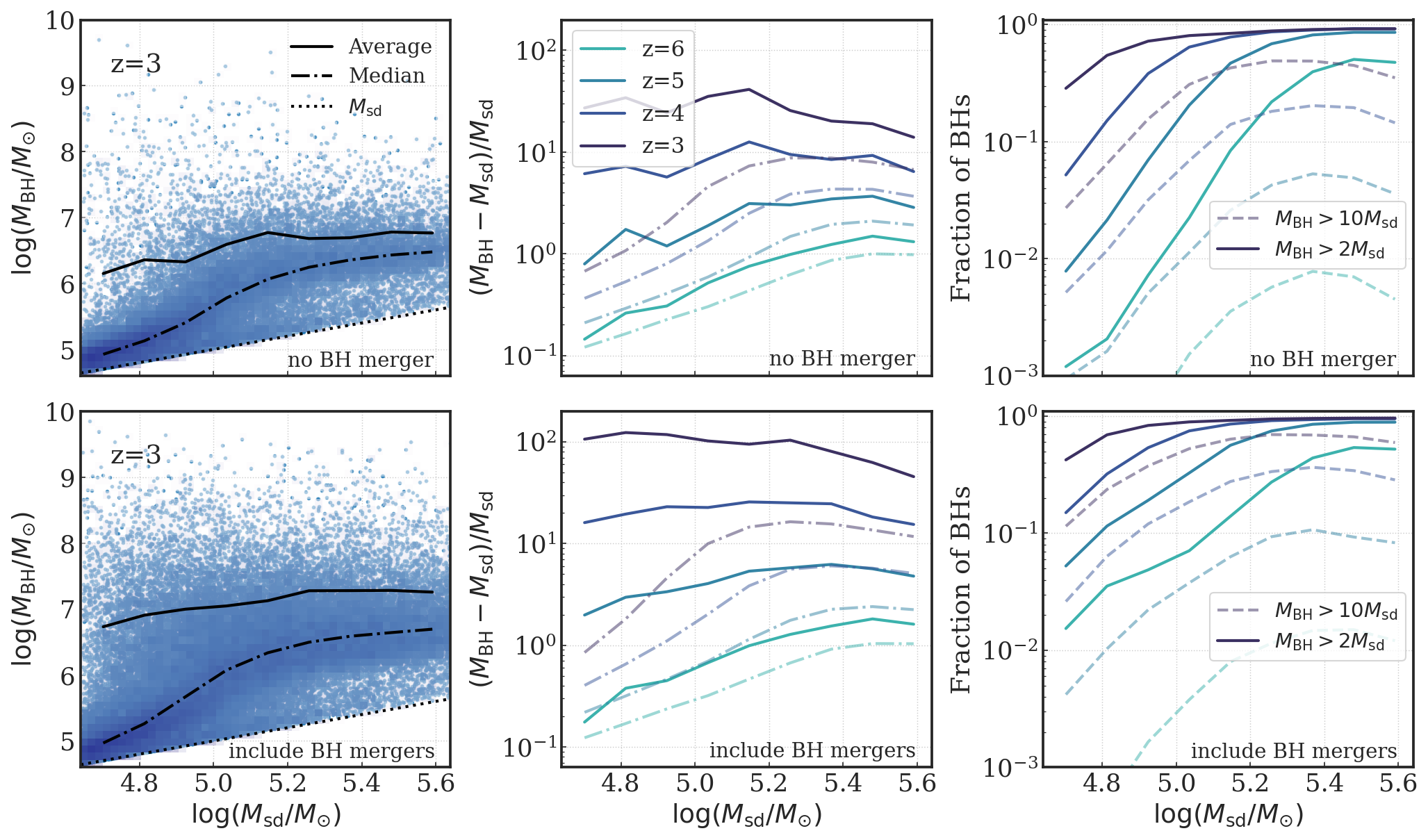}
  \caption{We trace the BHs seeded at $z>10$ down to lower redshifts and show their $M_{\rm BH}$ versus seed mass ($\msd$) at $z=3$ in the leftmost column. The black solid and dashed-dotted lines in the left column show the average and median $\mbh$ at each $\msd$ bin. 
  The middle column displays the mean and median $\mbh/\msd$ values for each $\msd$ bin from $z=6$ to $z=3$, represented by the solid and dashed-dotted lines with colours representing different redshifts.
  The right column gives the fraction of BHs in each $\msd$ bin that has grown beyond 2 times (solid lines) and 10 times (dashed lines) the initial $\msd$ at each redshift.
  The upper panels only include the BHs that have not undergone merger events.
  The lower panels include the BHs that have merged with other smaller mass BHs (i.e. being the more massive progenitor).}
  \label{fig:Mseed}
\end{figure*}

\begin{figure}
\centering
  \includegraphics[width=0.9\columnwidth]{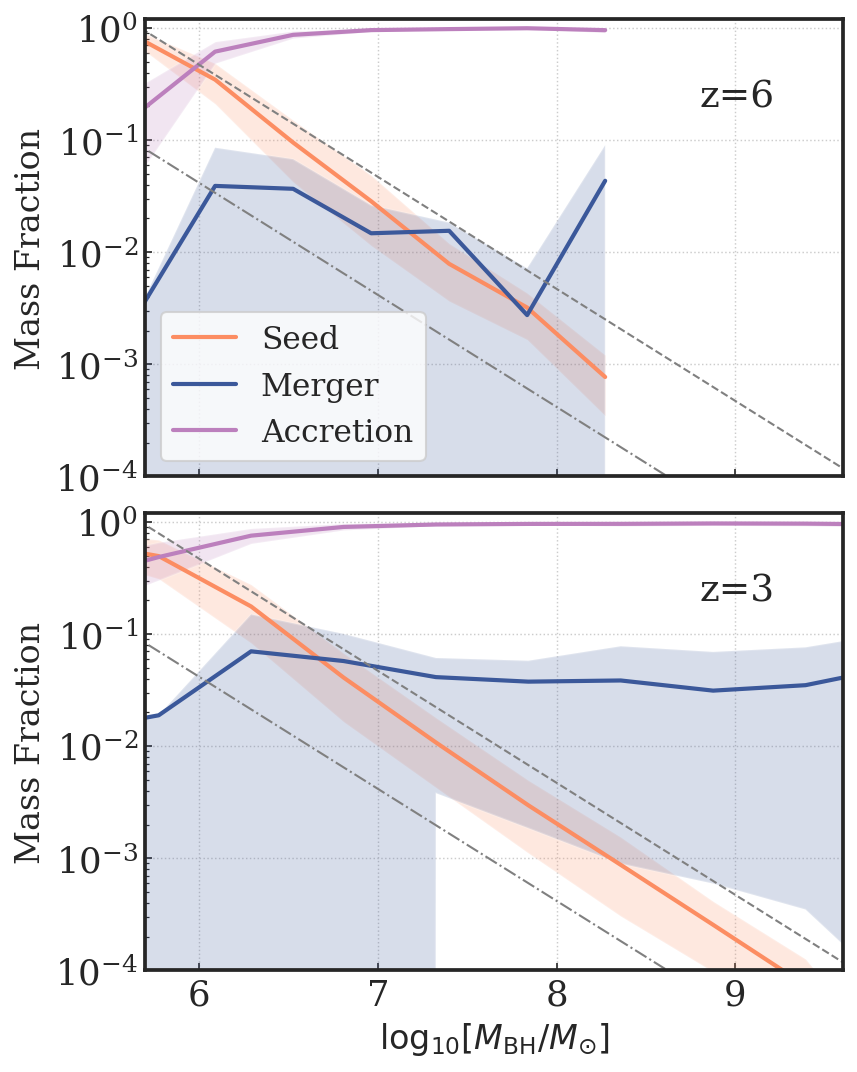}
  \caption{Fraction of BH mass contributed by the seed mass, $\msd$, by accretion and by mergers (swallowing other BHs) as a function of $\mbh$ at $z=6$ (top panel) and $z=3$ (bottom panel). For each BH that has undergone a merger, we use the $\msd$ from the more massive progenitor and sum the mass of the merged BHs, denoting that to be the mass contribution from mergers.
  The solid lines show the mean value for each $\mbh$ bin, with the shaded regions covering the range corresponding to the 16th and 84th percentiles. The grey dotted-dashed and dashed lines mark the range  $M_{\rm sd,min}/\mbh$ and $M_{\rm sd,max}/\mbh$ which are the lower and upper limits of the possible $\msd$ fraction.}
  \label{fig:BH-mass-inventory}
\end{figure}

\subsection{Effect of the BH seed mass}
\label{subsection:BH-seed}

Since the Bondi-Hoyle prescription for BH accretion depends on the BH mass with $\dot{M}_{\rm BH} \propto \mbh ^2$ (c.f. Eq.~\ref{equation:Bondi}), one would expect the early growth of BHs to be sensitive to the initial $\msd$. This is because in the early phase of BH growth the AGN feedback does not have a large impact on the local environment and the BH accretion is dominated by the $\mbh ^2$ term, where a larger $\msd$ would cause $\mbh$ to build up more rapidly.

In \asterix, BHs are seeded with $\msd$ stochastically drawn from a power-law distribution within the mass range from $M_{\rm sd,min} = 4.4 \times 10^4 \msun$ ($3 \times 10^4 \hmsun$) to $M_{\rm sd,max} = 4.4 \times 10^5 \msun$, which is independent of the host environment at the seeding time. 
This allows us to statistically investigate how the growth of a BH is dependent on its seeding mass.

To this end, we select a batch of BHs that are seeded in a given time interval, trace them to a later time, and show how their final $M_{\rm BH}$ are dependent on their initial $\msd$.
In order to allow the BHs enough time evolution, we select the BHs seeded at high redshift between $10 < z < 15$  (about $8.6\times10^4$ BHs in total), and trace them to lower redshifts from $z=6$ to $z=3$.
To take into account the fact that BHs also grow by merging with other BHs (with a different seed mass), we split the analysis into two parts. 
First, we exclude the BHs that have undergone merger events to cleanly diagnose how  $M_{\rm BH}$ gained by the gas accretion is affected by original seed mass. Second, we include the BHs that have merged with other smaller mass BHs to address the possible bias introduced by selecting the no-merger BHs.

The upper panels of Figure~\ref{fig:Mseed} show the result of the $\mbh - \msd$ relation for BHs that are seeded at $10 < z < 15$ and have not undergone a merger before the redshift inspected. 
The blue points in the left panel give the resultant $\mbh$ at $z=3$ versus their original $\msd$, with the black solid and dashed-dotted line giving the running mean and median $\mbh$ for each $\msd$ bin. 
The middle panel shows the statistics of the mean and median $\mbh$/$\msd$ in each $\msd$ bin from $z=6$ to $z=3$ to show time evolution.
The right panel shows the fraction of BHs in each $\msd$ bin that have grown beyond their initial $\msd$ at each redshift, with the solid and dashed lines representing $\mbh > 2\times \msd$ and  $\mbh > 10\times \msd$ respectively.

We can see that the initial $\msd$ has a clear imprint on the final $\mbh$, even after time evolution from $z=10$ to $z=3$.
The median $\mbh - \msd$ relation at $z=3$ displays a strong positive correlation especially at the lower $\msd$ end, indicating that for the BH population from the smallest $\msd$ bin, their BH accretion in general lags behind at the early stages and has not yet caught up even by $z=3$. 
The right panel tells that BHs with the smallest $\msd$ have a significantly lower chance to grow beyond the seed mass compared to BHs with more massive $\msd$: less than $50\%$ of BHs from the smallest $\msd$ bin have grown to $\mbh > 2\times \msd$ at $z=3$.

However, it is important to note that these results are not showing that $\mbh$ is predominantly determined by the $\msd$.
We can see from the left panel that the BHs from the lowest $\msd$ bin can still reach the highest $\mbh$ range ($\mbh > 10^9 \msun$) through pure gas accretion, indicating that the local environment specific to each BH plays a significant role in BH growth regardless of $\msd$.

Moreover, the $\mbh$ difference caused by different initial $\msd$ values becomes narrower as time evolves, as the $\mbh$ becomes dominated by the accretion process determined by the local environment.
The evolution of the $\mbh$-$\msd$ relation shown in the right two panels becomes flatter, especially for $\msd > 10^5\msun$ when moving to lower redshift.
At $z=3$, the median $\mbh / \msd$ ratio, as well as the fraction of $\mbh$ growth is almost the same for BHs from $\msd \sim 10^5\msun$ and from the most massive $\msd$ bin. 
This indicates that for BHs seeded with $\msd>10^5\msun$ at $z>10$, the impact of $\msd$ on the $\mbh$ at $z=3$ has been virtually erased by accretion and its associated exponential growth.

Selecting BHs that have not undergone any mergers might artificially eliminate the BH population residing in massive galaxies that are also likely to be gas-rich.
Therefore in the lower panel of Figure~\ref{fig:Mseed} we repeat the same analysis including the growth of BHs due to mergers.
In this case, the $\msd$ (in the lower panel of Figure~\ref{fig:Mseed}) is that of the more massive BH progenitor in the merger history. 
The resultant $\mbh$-$\msd$ relation displays the same qualitative features as in the upper panels, but with a higher $\mbh$ than obtained in the upper panel.

Therefore, we can conclude that the initial $\msd$ has an impact on $\mbh$ growth by gas accretion (and additionally by BH mergers, when they occur), whereby BHs from the smallest $\msd$ bin have a lower chance to grow beyond the seed mass even with the long period of time evolution from $z>10$ to $z=3$.
However, the effect of the $\msd$ would be erased with time evolution, and the $\mbh$ is not dominated by the initial $\msd$. 
The most massive BHs can grow from the smallest initial $\msd$, as the local environment for each specific BH plays a crucial role in their evolutionary history.

\subsection{BH mass fraction by seed, merger and accretion}
\label{subsection:BH-inventory}

In Section~\ref{subsection:BH-seed}, we have seen that the BHs seeded from the smallest $\msd$ bin can still grow to $> 10^9 \msun$ by gas accretion only.
This indicates that gas accretion plays a predominant role in the mass assembly history of at least some BHs.
It is interesting to analyse and compare the contribution of BH mergers and gas accretion to the overall $\mbh$ distribution on a statistical basis and investigate its time evolution.

To this end, in Figure~\ref{fig:BH-mass-inventory} we present the mass fraction of $\mbh$ contributed by $\msd$ and by mergers as a function of $\mbh$ at two different redshifts, $z=6$ and $z=3$.
For each BH that has undergone mergers, we trace the more massive progenitors, use the $\msd$ from the massive progenitor and sum the mass of the merged BHs (calling this the mass contribution from mergers).
The orange line in Figure~\ref{fig:BH-mass-inventory} shows the mass fraction contributed by $\msd$. It lies between $M_{\rm sd,min}/\mbh$ and $M_{\rm sd,max}/\mbh$, as marked by the grey dashed lines. 
The purple and blue lines give the averaged mass fraction of $\mbh$ obtained by BH accretion and by swallowing other BHs.

BHs in the mass range $\mbh \lsim 10^6 \msun$ are dominated by the seed mass, with the $\msd$ fraction lying toward the upper limit of $M_{\rm sd,max}/\mbh$. 
This is because the low mass BHs are more likely to trace the initial seed mass and BHs in the $10^6 \msun$ regimes have a larger seed contribution compared to those with massive seeds. The $\msd$ fraction becomes less biased when moving to the more massive end because massive BHs gain most of their mass from the gas accretion which is determined by the local environment. Their mass, therefore, is less dependent on the initial $\msd$.

The averaged mass fraction of $\mbh$ from BH mergers has an almost flat distribution at $z=3$, with no strong dependence on the BH mass.
\yy{The strict BH merger criteria applied in \asterix allows BHs to have a wide range of merger histories and leads to a large scatter in the BH mass fraction contributed by the merger. The scatter of the BH merger component (blue shade) increases with $\mbh$ as more massive BHs can involve BH mergers with larger mass contrast.}

The BH merger mass fraction in the $\mbh > 10^7\msun$ regime at $z=3$ is slightly larger than that at $z=6$, as more merger events happen at lower redshift. 
However, the averaged contribution to $\mbh$ from BH mergers is less than 10\% for all mass bins and at both redshifts.
For the population of the most massive BHs with $\mbh>10^7\msun$, on average, more than 90\% of the BH mass is obtained through the gas accretion, with only a few per cent from the BH mergers and negligible contribution from the seed mass.
Therefore, we conclude that in \asterix, BH mergers are a subdominant factor in BH growth until at least $z=3$.

\yy{Simulations where BHs are centered using repositioning rather than dynamic friction merge BHs more aggressively and thus make different predictions. For example, Illustris-TNG \citep{Weinberger2018} predicts that the SMBHs with $\mbh > 10^{8.5} \msun$ grow most of their mass via mergers with lower mass BHs (though this claim is for BH population at $z=0$). A more thorough comparison of the BH mergers with other simulations will be presented in our follow-up work.}
\section{BH-Galaxy Relations}
\label{section4:BH-Galaxy}

\begin{figure*}
\centering
  \includegraphics[width=1.0\textwidth]{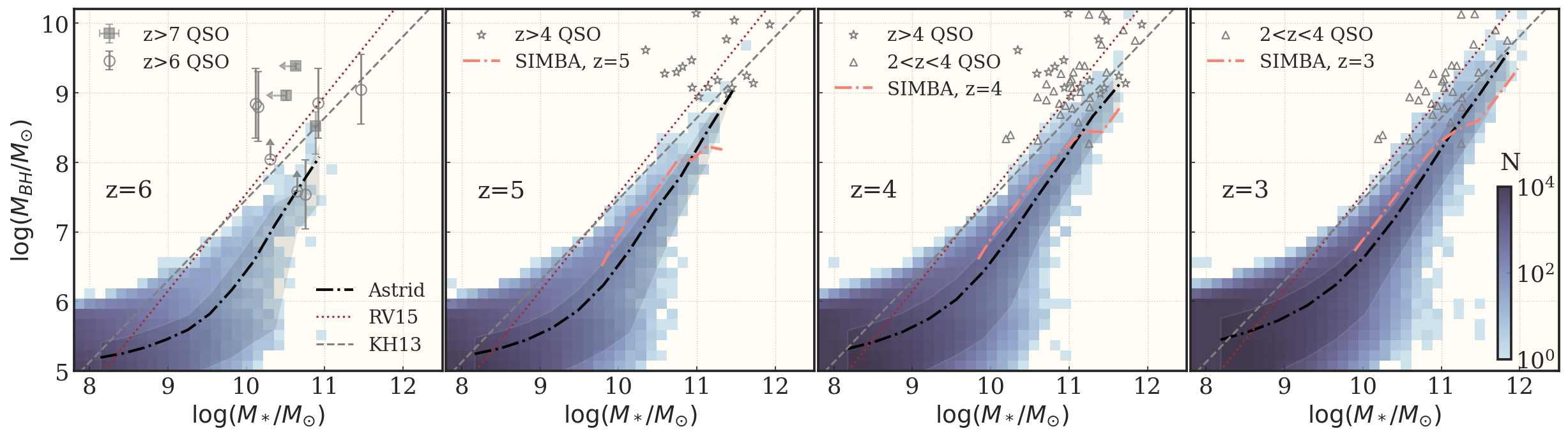}
  \caption{The scaling relations between $\mbh$ and galaxy mass $\mstar$ from $z=6$ to $z=3$ in \asterix, shown as 2D histograms. 
  The black dashed-dotted line shows the averaged $\mbh$ for each stellar mass bin, and the grey shaded area denotes the 16th and 84th percentiles. 
  The grey dashed line and brown dotted line show the fit results from \protect\citet{Kormendy2013,Reines2016}, which are based on $z\sim0$ AGN observations.
  The square data points are observations of three $z>7$ QSOs.
  The empty circles with error bars show the HSC observations of $z>6$ quasars from \protect\citet{Izumi2021}. 
  The star and triangle data points are $z>4$ and $2<z<4$ quasars collected from \protect\citet{Kormendy2013}.
  The red dashed-dotted line shows the $\mbh$ - $\mstar$ relation from the SIMBA simulation \protect\citet{Thomas2019-simba}.}
  \label{fig:scaling}
\end{figure*}

\begin{figure}
\centering
  \includegraphics[width=0.9\columnwidth]{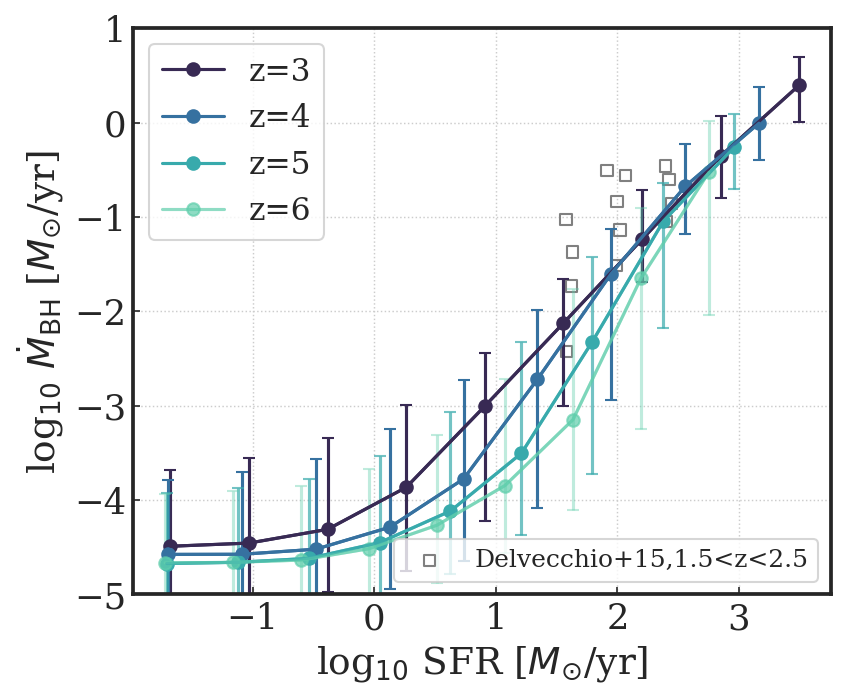}
  \caption{BH accretion rate $\dot{M}_{\rm BH}$ versus star formation rate (SFR) in the host galaxy, from $z=6$ to $z=3$. The solid lines show the running median with the error bars marking the 16-84th percentile in each SFR bin. The grey squares are from observations of a sample of $1.5<z<2.5$ star forming galaxies by \protect\citet{Delvecchio2015}.}
  \label{fig:bhmdot-sfr}
\end{figure}

\begin{figure}
\centering
  \includegraphics[width=0.9\columnwidth]{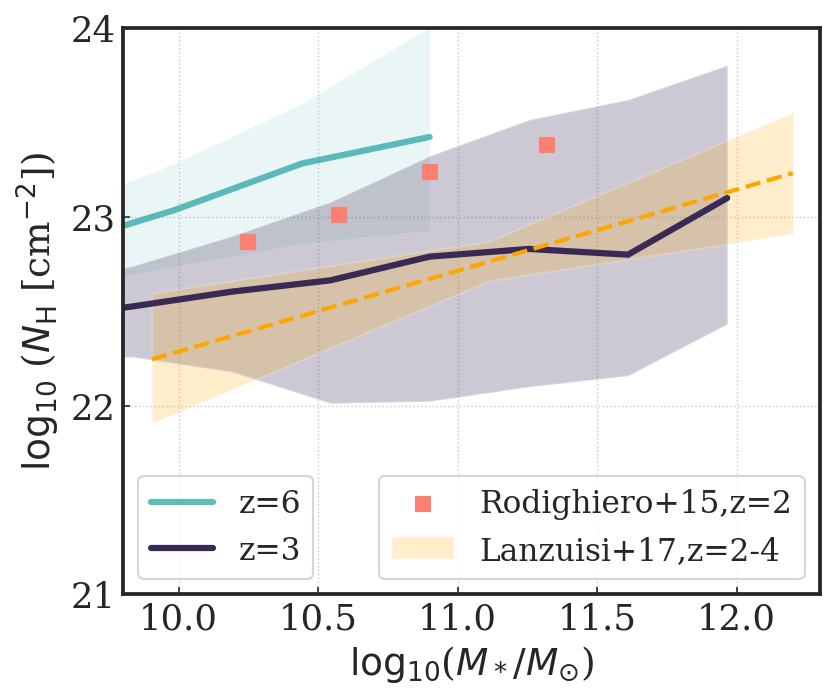}
  \caption{Gas column density of AGN as a function of the stellar mass in the host galaxy, including all AGN populations with $\Lx > 10^{42.5}$ erg s$^{-1}$ at $z=6$ and $z=3$. For each AGN, we calculate $\NH$ along 48 random lines of sight.  The solid black line shows the median of $\NH$, and the shaded area covers the 16-84th percentile of the $\NH$ distribution for all sightlines in each $\mstar$ bin. The red squares show observational results for $z\sim2$ AGN hosts from \protect\citet{Rodighiero2015}. The orange area is the linear regression of $\NH$ and $\mstar$ reported by \protect\citet{Lanzuisi2017} based on the $2<z<4$ AGN population. }
  \label{fig:NH-mstar}
\end{figure}

\begin{figure*}
\centering
  \includegraphics[width=1.0\textwidth]{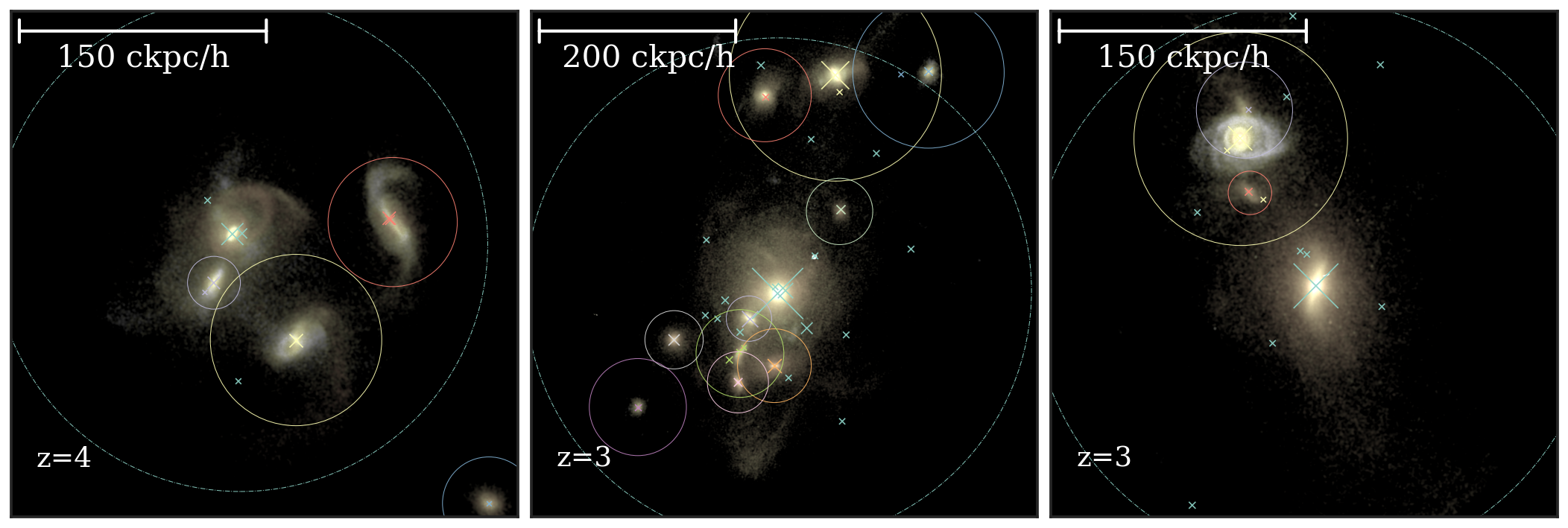}
  \includegraphics[width=1.0\textwidth]{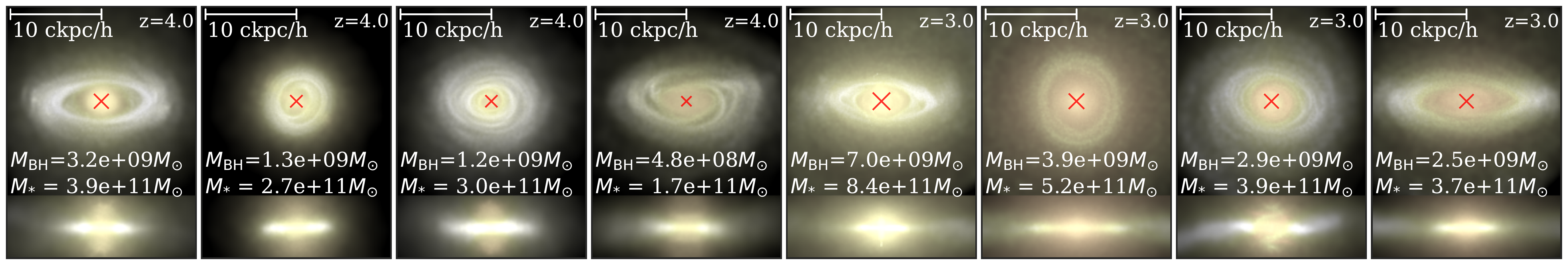}
  \caption{\textit{Top panels}: Visualizations of some $z=4$ and $z=3$ galaxies associated with different subgroups residing in the same FOF group. The coloured circles are the virial radii of the subgroups. The dashed circles correspond to the central galaxy (the most massive subgroup), where the circle radius is re-scaled by $0.5 $ to allow it to fit inside the figure. The coloured crosses mark the positions of the SMBHs associated with different subgroups, with the cross size scaled according to the BH mass.
  \textit{Bottom panel}: Samples of face-on and edge-on zoom-in views of the galaxy hosts of some massive BHs in \asterix at $z=4$ and $z=3$. The colour hue is determined by the age of the stars (with a red colour indicating older stars). All panels are centred on the position of the central SMBH. }
  \label{fig:Galaxy-image}
\end{figure*}

\begin{figure*}
\centering
  \includegraphics[width=0.95\textwidth]{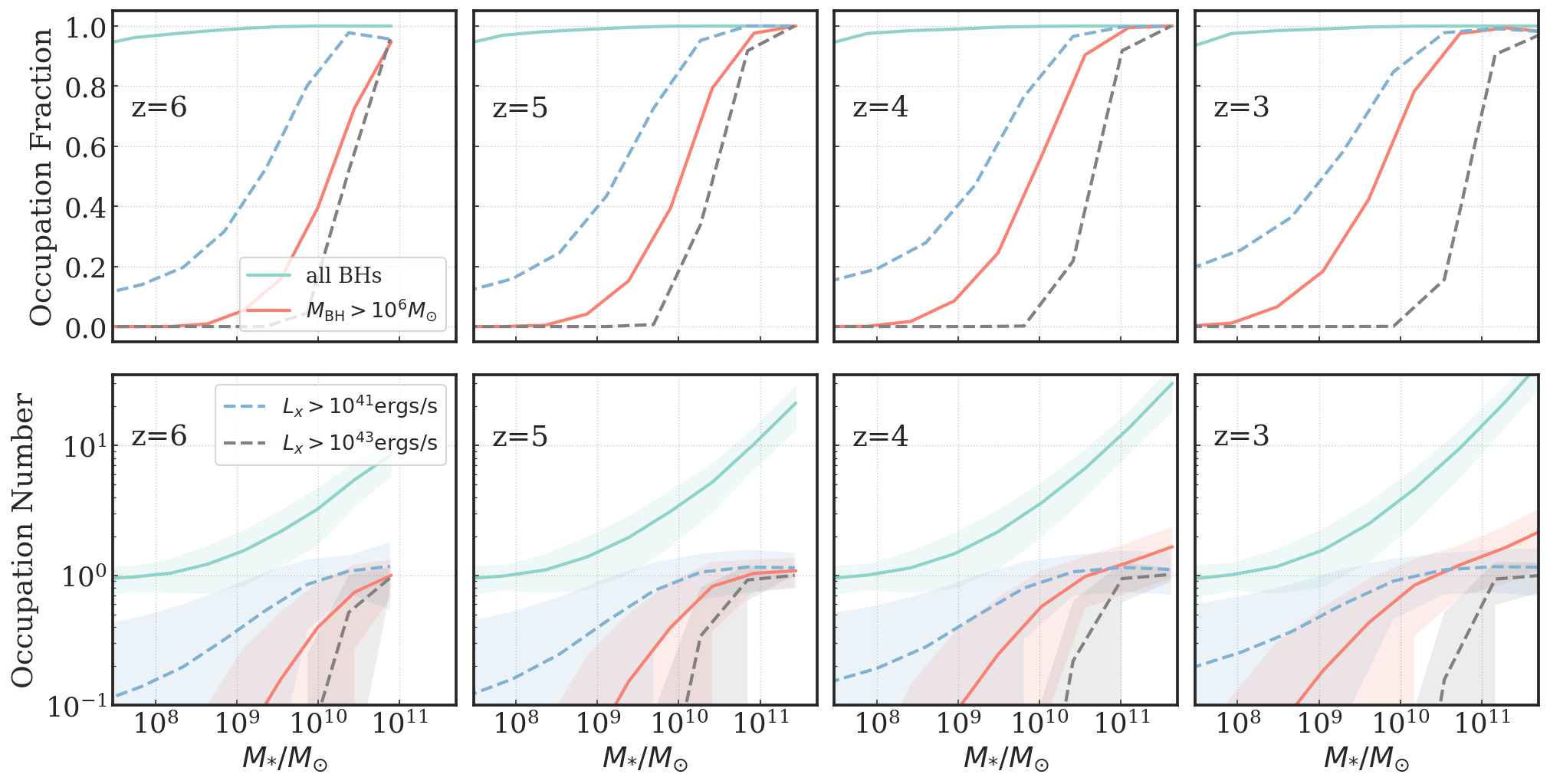}
  \caption{
  \textit{Upper panel:} The fraction of galaxies that at least host one BH in each stellar mass bin (on $y$-axis), from $z=6$ to $z=3$. The green lines include the entire BH population. The red lines are for the BH population with $\mbh > 10^6 \msun$. The blue and grey dashed lines include BH populations above different luminosity thresholds of $\Lx > 10^{41}$ erg s$^{-1}$ and $\Lx > 10^{43}$ erg s$^{-1}$ respectively. 
  \textit{Lower panel:} The occupation number of BHs as a function of the stellar mass. The coloured lines show the average number of BHs for galaxies residing in each stellar mass bin, with the same colour conventions as the upper panels. 
  The shaded area covers the 16-84th percentile.}
  \label{fig:Occupation} 
\end{figure*}

We identify galaxies and BHs in the subgroups by running \textsc{SUBFIND} \citep{Springel2001} on the FOF particle catalogue in post-processing. 
\yy{The statistics of galaxies, such as the stellar mass function and galaxy star formation rate, can be found in our companion paper \citep{Bird2021}.}
In this section, we investigate the relationships between BHs and their host galaxies.
We look into the scaling relation between $\mbh$ and $\mstar$ in \S~\ref{subsection: mbh-mstar} and the scaling relation between $\dot{M}_{\rm BH}$ and host star formation rate in \S~\ref{subsection:mdot-sfr}.
\S~\ref{subsection: NH-mstar} is focused on the relationship between AGN galactic scale obscuration and the host $\mstar$.
\S~\ref{subsection: BH-occupation} investigates the statistics of BH occupation in host galaxies. 

\subsection{The scaling relation between $M_{\rm{BH}}$ and $M_*$}
\label{subsection: mbh-mstar}

Scaling relations between central BH mass and host galaxy properties are of fundamental importance to studying BH and galaxy evolution through cosmic time.
In Figure~\ref{fig:scaling}, we show the scaling relations between $\mbh$ and $\mstar$ from $z=6$ to $z=3$ in \asterix.
Here we use the total stellar mass of the galaxy in the subhalo and do not apply bulge-disk decomposition to identify the galaxy bulge. 
We only consider the central galaxies in each FOF halo (the main subhalo in each FOF halo with the greatest stellar mass) and do not include satellite galaxies in this analysis. 
For each subhalo that contains multiple BHs, we select the most massive one. 
We have also repeated the analysis by selecting the most luminous BH for each subhalo and this does not lead to any significant difference.
The grey dashed and the brown dotted line shows the best-fitting relations from \cite{Kormendy2013,Reines2016} based on the AGN population in the nearby universe.
We also collect the data points from observations of high-$z$ QSOs. 
The squared data points are properties of three $z>7$ QSOs summarized in \cite{Izumi2019} and the circles with error bars show Subaru HSC observations of $z>6$ QSOs from a recent compilation by \cite{Izumi2021}. 
The observational stellar masses presented here are the inferred dynamical masses of the host galaxies.
The star and triangle-shaped data points are for $z>4$ and $2<z<4$ quasars collected from \cite{Kormendy2013}.

Although the box size of \asterix is relatively large, it is not large enough to produce any $10^8\msun$ BH at $z>7$ or $\sim 10^9 \msun$ BHs at $z>6$, and so we cannot directly compare with $z>6$ QSO observations. 
The most massive BHs ($\mbh>10^8\msun$) at $z=6$ reside in $\mstar \sim 10^{11}\msun$ galaxies, and minimal extrapolation of these \asterix results leads to qualitatively good agreement with the HSC observations. 
The BH population with mass $\mbh < 10^6 \msun$ is dominated by BHs which are at the inefficient initial growth phase close to the seed mass. There is a large variation with regard to the host galaxy mass, ranging from $\mstar = 10^8 \sim 10^{10} \msun$. 
At $\mbh > 10^6 \msun$, where BHs go beyond the seed mass, \asterix produces a clear positive correlation between the stellar mass of a galaxy and the mass of its central BH, in concordance with the expectation that galaxies and BHs grow commensurately in a globally averaged sense. 

Compared with fits to AGN populations in the local universe from \cite{Kormendy2013,Reines2016}, \asterix has $\mbh$ averaging about 0.5 dex lower for each $\mstar$ bin, with the discrepancy getting narrower at the more massive end with $\mstar > 10^{11} \msun$. 
The lower averaged $\mbh$ predicted by the simulation can be partially attributed to the population of small BHs that are difficult to obtain using observations.
Moreover, we note that this discrepancy is of the same order as likely systematic errors in the observations and that these systematics will be worse for low mass objects.

As a comparison with other simulations, we also plot using a pink dashed-dotted line the $\mbh - \mstar$ relation measured from the SIMBA simulation \citet{Dave2019-simba}.
We can see that even with different implementations of the BH accretion and feedback, \asterix and SIMBA produce $\mbh - \mstar$ relations in good agreement with each other.

\subsection{BH accretion and SFR relation}
\label{subsection:mdot-sfr}

Since active BH growth is induced by the same cold dense gas that fuels star formation, one expects that the star formation rate and BH accretion rate should be positively correlated.
This is supported by a range of AGN observations \citep[e.g.][]{Mullaney2012,Chen2013,Delvecchio2015,Lanzuisi2017} and is also established in many simulations \citep[e.g.][]{Thomas2019-simba,Ricarte2019}.

In Figure~\ref{fig:bhmdot-sfr}, we show the instantaneous $\dot{M}_{\rm BH}$ - SFR relationship from $z=6$ to $z=3$.
The solid lines are for the median $\dot{M}_{\rm BH}$ for each SFR bin, with the error bars showing the range corresponding to the 16-84th percentile.
The grey squares are observational samples collected from the work of \cite{Delvecchio2015} who combined IR band selected star-forming galaxies and X-ray detected AGNs over the redshift range covering $1.5<z<2.5$.

The SFR broadly trace the $\dot{M}_{\rm BH}$ for galaxies with $\mathrm{SFR} \gtrsim 1 \msun/{\rm yr}$, and show reasonable agreement at the order of magnitude level compared to observational data for active star-forming galaxies \citep{Delvecchio2015}.
The curves flatten at the low SFR end, as smaller galaxies with lower SFR host BHs near the seed mass, where $\dot{M}_{\rm BH}$ is dominated by $\mbh$ and is not on average coupled to SFR. 

The $\dot{M}_{\rm BH}$-SFR relation at different redshifts does not show obvious time evolution for active star-forming galaxies with $\mathrm{SFR} \gtrsim 100 \msun/{\rm yr}$.
For AGN hosts with SFR $< 100 \msun/{\rm yr}$, the $\dot{M}_{\rm BH}$-SFR relation exhibits substantial scatter, and the median $\dot{M}_{\rm BH}$ for the same SFR decreases when going to higher redshift. 
This is because, for smaller hosts, a larger fraction of BHs have not built up the mass to reach the self-regulated regime and $\dot{M}_{\rm BH}$ is dominated by the initial Bondi accretion that is sensitive to $\mbh$, where $\mbh$ is smaller at higher redshift. We note that the $\mbh - \mstar$ relation shows a similar trend with redshift for smaller galaxies. 


\subsection{Relationship between $\NH$ and host galaxy properties}
\label{subsection: NH-mstar}

As another probe of the connection or feedback between SMBH growth and galaxy build-up, observations are used to investigate the relationship between AGN gas obscuration and the host galaxy mass, with some finding a positive correlation between $N_{\rm H}$ and $M_*$ \citep[e.g.][]{Rodighiero2015,Buchner2017a,Lanzuisi2017}.

In Figure~\ref{fig:NH-mstar}, we show the simulation predictions for the $\NH$ - $\mstar$ relation.
We investigate the $\Lx > 10^{42.5}$ erg s$^{-1}$ AGN populations and their host galaxies at $z=6$ and $z=3$.
The black solid line surrounded by a shaded area shows the running median and 16-84th percentiles of the $\NH$ distribution including all sightlines for AGN lying in each $\mstar$ bin.
For comparison, the red squares in Figure~\ref{fig:NH-mstar} show observational results from \cite{Rodighiero2015} for a sample of $z \sim 2$ AGN hosts in the COSMOS field. 
The purple shaded region is the linear fit of $\NH$ to $\mstar$ for the $2<z<4$ QSO population measured by \cite{Lanzuisi2017}, based on a sample of X-ray detected AGN and their far-UV detected host galaxies.

\asterix predicts an overall higher normalization for $\NH$ at $z=6$ compared to $z=3$, as the galactic obscuration at higher redshift is larger. 
Considering the large variations brought about by different AGN sightlines, \asterix predicts an $\NH$-$\mstar$ relation broadly consistent with observations.
We remind the reader that $\NH$ calculated in the simulation is the galactic scale gas obscuration, which does not include resolving the nuclear torus, and is therefore a lower limit to the obscuration that would be probed by AGN observations.
The consistency with observations implies that a significant fraction of the obscuration observed in AGN is due to galaxy-scale gas in the host \citep[as also claimed in, e.g.][]{Buchner2017a}. 

\asterix predicts a weak positive correlation between $\NH$ and $\mstar$ at both $z=6$ and $z=3$, implying that as galaxy mass increases there are higher chances to have an additional component to the amount of gas along the line of sight towards the AGN.
However, we also note that the variation predicted by the simulation is large. 
At $z=3$, the $1\sigma$ variation for massive galaxies with $\mstar>10^{11}\msun$ is about 1.5 dex, ranging from $\NH = 10^{22} \rm{cm}^{-2}$ to $10^{23.5} \rm{cm}^{-2}$.
As discussed in our previous work, \cite{Ni2020}, simulations predict a large angular variation of $\NH$ values along different AGN sightlines, as the AGN feedback modulates the surrounding gas density and launches gas outflows that lead to windows of low obscuration within even very massive galaxy hosts. 
We therefore do not see a tight power-law relation between $\NH$ and $\mstar$ due to this large variation between different sightlines.

\subsection{BH occupation in galaxies}
\label{subsection: BH-occupation}
There is compelling evidence that almost every massive galaxy in the nearby universe contains an SMBH at its centre. However, this may not be true for fainter AGN in low mass galaxies
and the simulation will help us determine what to expect and why.
The fraction of galaxies hosting AGN as a function of galaxy mass and other properties also provide useful information about the growth and fueling of the first BHs.
In addition, it is interesting to investigate the statistics of BH occupation numbers in galaxies using a simulation that includes physically-based modelling of BH mergers and how they sink to the centre of their hosts.
In \asterix, BHs do not merge immediately following the mergers of host galaxies, but instead, their relative kinetic energies are dissipated by the force of dynamical friction and they become gravitationally bound.
As a consequence, large galaxies in \asterix will in general host multiple BHs. 
In this section, we show statistics of the BH occupation in galaxies.

Figure~\ref{fig:Galaxy-image} provides a visualization of the galaxy hosts for BHs.
The upper panel illustrates how BHs reside in galaxies associated with different subgroups, all from the same large FOF group.
We show some examples of large systems at $z=4$ and $z=3$. 
Different subgroups are identified by coloured circles that correspond to the virial radius $R_{200}$ of that subgroup.
The dashed circles identify the central galaxy (associated with the most massive subgroup). 
The coloured crosses mark the positions of the BHs corresponding to different subgroups, with the cross size scaled by the BH mass.
The lower panels of Figure~\ref{fig:Galaxy-image} show face-on and edge-on zoom-in views of the galaxy hosts for a selection of massive BHs at $z=4$ and $z=3$. The colour of the galaxies is determined by the age of the stars, with redder colours representing older stellar populations.

The upper panel of Figure~\ref{fig:Galaxy-image} illustrates how massive galaxies can host multiple BHs. 
In Figure~\ref{fig:Occupation}, we present quantitative information on BH occupation as a function of the stellar mass in the host galaxy.
The upper panel shows the fraction of galaxies that contain at least one BH (occupation fraction) in each stellar mass bin, from $z=6$ to $z=3$.
The green line includes all BHs, while the red line includes only the BHs that have grown beyond the seed mass regime with $\mbh > 10^6 \msun$.

We also apply luminosity thresholds to predict BH occupation for bright AGNs and also for fainter AGNs that could be detected by the future X-ray observatory Lynx.
The Lynx flux limit of $1 \times 10^{-19}$ ergs cm$^{-2}$ $s^{-1}$ in $2-10$ keV band \citep[see, e.g.][]{Lynx2018} 
allows it to probe an AGN population with $\Lx \gtrsim 10^{40-41}$ erg s$^{-1}$ at redshifts $z=7\sim3$.
The blue and grey dashed lines represent the occupation number of BHs with $\Lx > 10^{41}$ erg s$^{-1}$ and $\Lx > 10^{43}$ erg s$^{-1}$ respectively.

The lower panel of Figure~\ref{fig:Occupation} displays the BH occupation number as a function of galaxy stellar mass, showing the predicted number of BHs for galaxies with given $\mstar$. The coloured lines give the mean occupation number and the shaded area covers the 16-84th percentiles for all galaxies in each $\mstar$ bin.

Across all redshifts from $z=6$ to $z=3$, galaxies with $\mstar = 10^9 - 10^{10} \msun$ begin hosting BHs that have grown beyond the seed mass or are relatively active with luminosities  $\Lx > 10^{41}$ erg s$^{-1}$.  
\yy{This is qualitatively consistent with the findings of many other simulations such as Illustris, Illustris-TNG, Horizon-AGN and SIMBA, where $\mstar \sim 10^{10} \msun$ galaxies have a probability of > 80 per cent to host an efficient accretor at $z=3\sim4$ (with threshold $L_{\rm bol} > 10^{43}$ erg s$^{-1}$, see \cite{Habouzit2022} for a comprehensive review).}
On the other hand, most galaxies below $10^9 \msun$ contain only BHs near the seed mass. 
The deficiency of BH growth in low mass galaxies could be explained by the lack of cold dense gas that fuels BH accretion in the shallower potential well of smaller galaxies.
Moreover, multiple simulation studies have found that BH growth in low mass galaxies with $\mstar < 10^9 \msun$ is regulated or stunted by supernovae (SN) feedback so that low mass galaxies generally have a lower probability of hosting a BH \citep[see, e.g.,][]{Dubois2015,Habouzit2017}.

The green line in the lower panel of Figure~\ref{fig:Occupation} shows that the mean occupation number of BHs quickly increases with stellar mass. 
A similar feature is found in other cosmological simulations where BHs are permitted to evolve dynamically without being fixed to halo centres \citep[e.g.][]{Volonteri2016Horizon-AGN,Tremmel2018b-wanderBH,Ricarte2021}. 
In particular, \cite{Ricarte2021} shows that the occupation number of wandering BHs scales with halo or galaxy mass as a power-law relation. 

We note that the occupation number of more massive and more luminous BHs becomes flattened for large galaxies.
At $z=3$, the most massive $\mstar>10^{11}\msun$ galaxies on average host 2 BHs with $\mbh > 10^6 \msun$, and about $1 \sim 2$ actively accreting BHs with $\Lx > 10^{41}$ erg s$^{-1}$.
\yy{The slower increase in luminous AGN with galaxy mass can be explained by efficient AGN feedback in massive galaxies which self-regulates the BHs and so suppresses the number of rapid accretors. 
A similar feature is also seen in some other simulations like Illustris-TNG and SIMBA that apply very strong AGN feedback for massive BHs \citep[see, e.g.][for a review]{Habouzit2021, Habouzit2022}. In those cases, they find that massive galaxies with $\mstar \sim 10^{11}\msun$ are even less likely to host AGN with rapid accretion at $z \leq 3$ compared to galaxies with $\mstar \sim 10^{10}\msun$.}
Another reason of the flat feature of the BH occupation number in massive galaxies is likely caused by BH mergers. 
As we will show in Section~\ref{section5:BH-merge}, more massive galaxies, in general, have undergone more BH mergers. Also, more massive BHs on average have a higher merger rate compared to seed mass BHs.
Therefore, even if massive galaxies contain several unmerged seed mass BHs, the average number of central actively accreting BHs remains $1 \sim 2$.


\begin{figure*}
\centering
  \includegraphics[width=1.0\textwidth]{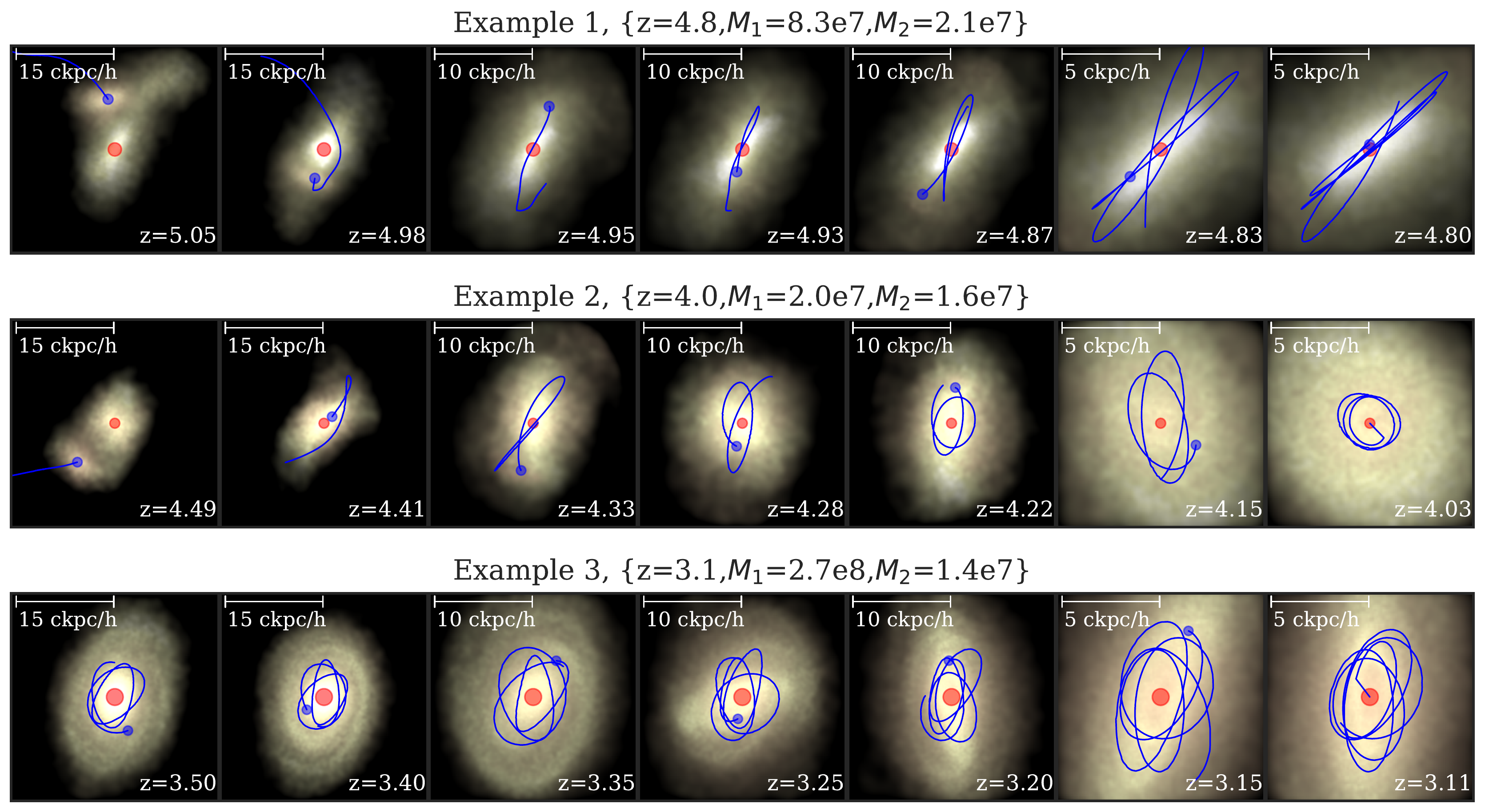}
  \includegraphics[width=1.0\textwidth]{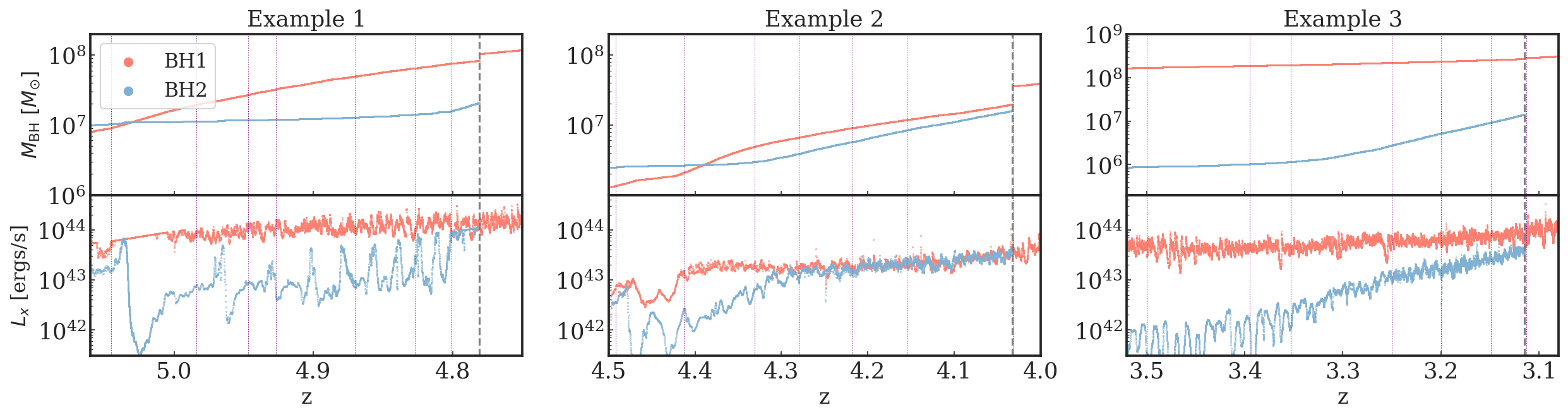}
  \caption{Three examples of merging galaxies and the evolution of their central BHs. 
  \textit{Top panels:} Visualization of galaxies and their corresponding BH orbital trajectories. The three rows show the evolution of three merger examples respectively. The background is the galaxy density field hued by the age of the star, with a redder shade representing older galaxies. All panels are centred at the position of the primary BH (the more massive object at merger time). We keep zooming into smaller (down to 5 ckpc$^3$) regions when approaching the merger time to better illustrate the morphological details of the host galaxies.
  \textit{Lower panels:} The detailed evolutionary history of the BH mass and luminosity for the three merger systems shown above, with red and blue lines representing the primary and the secondary black hole (BH1 and BH2) respectively. The purple vertical lines mark the time of the snapshots shown in the top panels. The grey dashed line marks when the merger event happens. The light curves of the merging systems shown here are all above the Lynx satellite detection limit.}
  \label{fig:Merge-examples}
\end{figure*}

\begin{figure*}
\centering
  \includegraphics[width=1.0\textwidth]{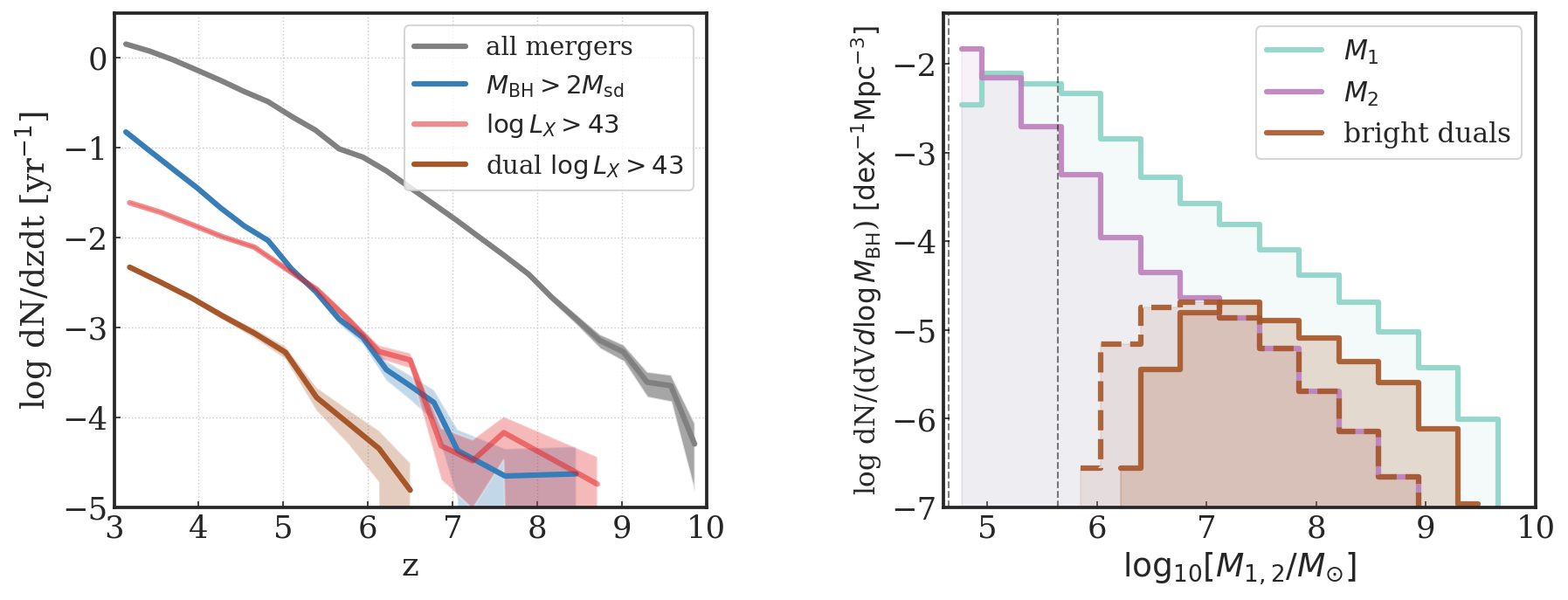}
  \caption{\textit{First panel}: The global BH merger rate from $z=10$ to $z=3$. The grey line includes all  BH mergers. The blue line covers only the merger population where both BHs have grown beyond their seed mass with $\mbh > 2\times \msd$ at the moment of the merger. The red line gives the merger rate for binaries where at least one BH has luminosity $\Lx > 10^{43}$ erg s$^{-1}$ at the time of the merger. 
  Finally, the brown line shows merger events where both BHs have ever had $\Lx > 10^{43}$ erg s$^{-1}$ (simultaneously) when at a separation $dr < 10 \hkpc$ before the merger.
  \textit{Second panel}: BH mass function of merger BHs, giving the $\mbh$ distribution of $M_1$ (green line) and $M_2$ (purple line) for all the BH merger events happened at $z>3$, normalized by the box volume and $\mbh$ bin. The brown solid and dashed lines show the distribution of $M_1$ and $M_2$ for bright dual AGN mergers with both $\Lx > 10^{43}$ erg s$^{-1}$, as indicated by the brown line in the left panel.
  }
  \label{fig:BH-merger-history}
\end{figure*}

\begin{figure}
\centering
  \includegraphics[width=0.95\columnwidth]{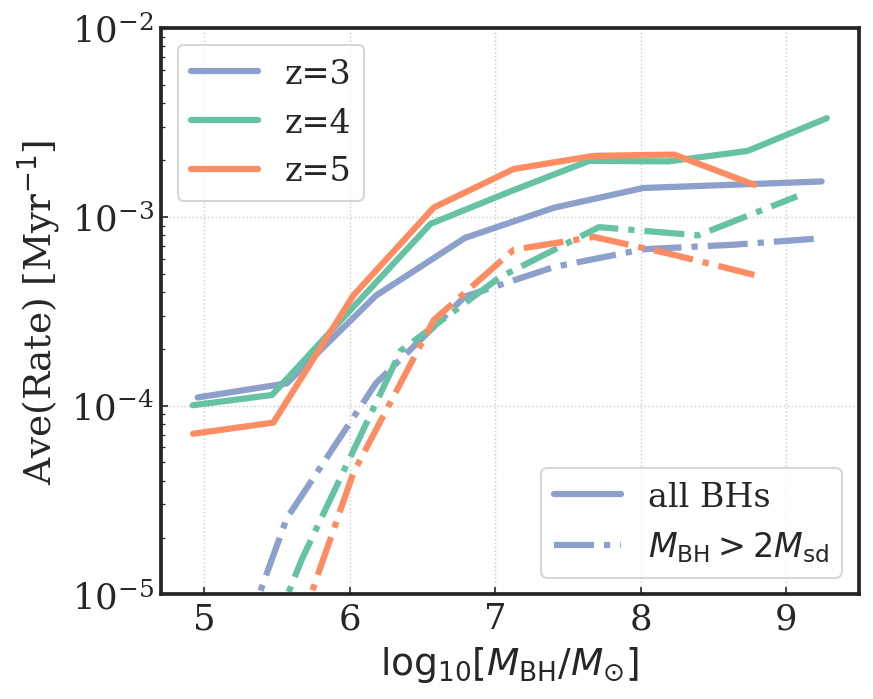}
  \caption{The averaged merger rate for BHs in a given $\mbh$ bin, showing the rate at which an individual BH of a given mass would be expected to undergo mergers. 
  Orange, green and blue lines show results at redshifts $z=5$, $z=4$ and $z=3$ respectively. 
  The solid lines include all BH mergers, and the dashed-dotted lines include only the non-seed BH merger events where both BHs have $\mbh > 2\times \msd$ at the moment of merger.}
  \label{fig:BH-merger-mbh}
\end{figure}

\section{BH Mergers}
\label{section5:BH-merge}

In \asterix, we use a sub-grid model to follow the dynamical evolution of SMBHs down to the spatial resolution corresponding to the gravitational softening length $\epsilon_{\rm g} = 1.5 \hkpc$, and numerically merge the bound BH pairs when they reach separation $|\bf{\Delta r}|$ $< 2 \epsilon_{\rm g}$.
In reality, BH binaries should take a significant time to in-spiral towards each other on scales below $1 \hkpc$ before they actually merge. 
However, in cosmological simulations, we cannot resolve the dynamical friction that dominates BH trajectories on sub-kpc scales and therefore we are unable to model this process in a self-consistent way. 
In follow-up work, \cite{Chen2021b}, we will apply a sub-grid model to follow the process of BH merging below the spatial resolution of \asterix using postprocessing techniques and make forecasts for upcoming observations of BH binaries and gravitational waves.

For each BH merger event, we denote the BH with the larger mass at the time of merger as the primary BH and the BH with the smaller mass as the secondary. We use $M_1$ and $M_2$ to represent their corresponding $\mbh$ at the merger time. Over the course of the simulation (down to $z=3$) we find a total of $4.5 \times 10^5$ BH merger events.

The current section presents the first-order results of the BH mergers in \asterix as determined by a physically motivated and self-consistent treatment of the dynamical friction and BH mergers in the context of cosmological simulations.
We first show some examples of BH merger events in \S~\ref{subsection: bh-merger-example}.
In \S~\ref{subsection: bh-merger-stats}, we present the global statistics that describe BH mergers in \asterix,
and in
\S~\ref{subsection: bh-merger-host} we investigate how the BH merger population is distributed in different galaxies. 

\subsection{Examples of merger events}
\label{subsection: bh-merger-example}

We begin with some illustrations of the BH mergers in \asterix. 
Figure~\ref{fig:Merge-examples} shows three examples of BH merger events, happening at different redshifts and with different $\mbh$.
The top three panels plot the trajectories of the merging BHs together with the evolution of their host galaxies shown in the background.
All the snapshots are centred on the position of the primary black hole (BH1, red point). The blue lines show the trajectories of the secondary black hole (BH2, the blue point) in the reference frame of BH1 over a time period with a redshift interval  $dz = 0.2$. 
The smooth trajectory of the merging BHs is brought about by the well-defined BH particle velocity.

The lower panels of Figure~\ref{fig:Merge-examples} show the evolutionary histories of the BH masses and luminosities for the merger examples, with the red and blue lines representing BH1 and BH2 separately. We give the X-ray band luminosity  (c.f. Eq.~\ref{equation:LbolLx}) modelled from the BH accretion rate. 
Note that the merging BHs in these three examples are well above the Lynx detection limit of $\Lx \gtrsim 10^{40-41}$ erg s$^{-1}$.

In both examples 1 and 2, the BH merger takes place not long after the encounter or merger of their host galaxies (as shown in the first panel).
For the first few close encounters, the BH binaries have large relative velocities that do not satisfy the merger criteria. 
With their kinetic energy gradually dissipated by dynamical friction, the binary systems become gravitationally bound and the BHs (numerically) merge at the last encounter. 
Example 1 shows a relatively eccentric trajectory before the merger, while in Example 2 the BH relative orbit is more circular.
The masses of the BHs in the Example 2 binary are very close to each other with $q \equiv M_2/M_1 = 0.8$.
They grow commensurately during the merger process, as also shown by their respective light curves $\Lx$ in the lower panel.
In both examples 1 and 2, the luminosities of the binary BHs are similar at the time of the merger, as the two BHs are close in $\mbh$ and are embedded in similar gas environments by the time they reach the merger criterion $|\bf{\Delta r}|$ $< 3 \hkpc$.

The third panel shows an example where the merger happens a long time after the first encounter between the host galaxies. In this case, BH2 lingers in the host galaxy before eventually sinking to the galaxy centre and merging with BH1. 
In this case the two BHs have a larger mass difference, with $M_2 \sim 10^6 \msun$ and $M_2 \sim 10^8 \msun$ at $z=3.5$. BH2 accretes strongly, reaching a larger mass of $10^7 \msun$ at $z=3.1$ during the process of sinking to the galaxy centre. It has slightly lower final luminosity compared to BH1, as $M_2$ is still an order of magnitude less than $M_1$.

\subsection{Global Statistics of BH mergers}
\label{subsection: bh-merger-stats}
\subsubsection{Global BH mergers}

We start by examining the global statistics of BH merger rates with respect to redshift and BH mass.
The first panel of Figure~\ref{fig:BH-merger-history} gives the BH merger rate as a function of redshift.
Here the merger rate is defined as the rate of GW signals that will reach the Earth from BH mergers happening at a given redshift, obtained by integrating the number of mergers over the redshift range and incorporating the cosmic volume at that redshift bin:
\begin{equation}
    \frac{{\rm d}N}{{\rm d}z\,{\rm d}t} = \frac{1}{z_2-z_1} \int_{z_1}^{z_2} \frac{{\rm d}^2 n(z)}{{\rm d}z\,{\rm d}V_c} \frac{{\rm d}z}{{\rm d}t} \frac{{\rm d}V_c}{{\rm d}z} \frac{{\rm d}z}{1+z}\,
\end{equation}
here $V_c$ is the comoving volume of the redshift bin and $n(z)$ is the number of mergers at a given redshift. 

The second panel of Figure~\ref{fig:BH-merger-history} shows the distribution of $M_1$ and $M_2$ values for all the BH merger events in green and purple histograms respectively. The $y$ axis gives the number of BHs that have encountered mergers at $z>3$ in the simulation, normalized by the volume and $\mbh$ bin.
The merger events cover $\mbh$ over the entire mass band, with  the $M_2$ distribution trending systematically smaller, as expected given that $M_2$ is the mass of the secondary BH in each merger event.

Most of the BH mergers involve BHs within the seed mass range, as small mass BHs are the dominant population. 
The grey line in the left panel of Figure~\ref{fig:BH-merger-history} shows the overall BH merger rate from $z=10$ to $z=3$.
We also separately examine the component of mergers that do not involve seed mass BHs.
In this work, we define the non-seed mass mergers as merger events where both BHs have grown beyond two times  their original seed mass $\mbh > 2\times \msd$ at the moment of the merger. 
There are in total $3.2 \times 10^4$ such non-seed mass mergers occurring between the start of the simulation and $z=3$.
The blue line shows the merger rate history of these non-seed mass mergers, which is $1 \sim 2$ dex lower than the global merger rate.

We also investigate the merger of bright AGNs that could be detected through their electromagnetic signatures.
In particular, we are interested in the case where one or two of the BHs are bright AGN before the merger. 
The red line shows the mergers for binaries where at least one BH has X-ray band luminosity $\Lx > 10^{43}$ erg s$^{-1}$ at the time of the merger. In total there are $8 \times 10^3$ such mergers.
We are also interested in bright dual AGN where both of the merging BHs are luminous before the merger. 
As a brown line, we show the rate of merger events where both BHs ever have $\Lx > 10^{43}$ erg s$^{-1}$ simultaneously when below separation $dr < 10 \hkpc$ before the merger take place. 
\yy{We note that the dual AGN population discussed here is a subset of the bright AGN pairs that would undergo merger at $z>3$ in \asterix, and therefore do not represent the overall dual AGN population at a given redshift.}
There are in total $\sim 1\times10^3$ such bright dual AGN mergers at $z>3$.
These bright dual AGNs are typically massive BHs, with typical mass $10^{7-8} \msun$, as shown by the mass distribution (brown lines) in the right panel.

\subsubsection{BH Merger rates as a function of $\mbh$}

Detailed information on each of the half-million BH mergers in the simulation is available to us, allowing us to study different aspects of their demographics.
We next turn to the masses of merging BHs. 
In Figure~\ref{fig:BH-merger-history}, where in the right panel we can see that most of the BH mergers are composed of small mass BHs. 
This is because the overall $\mbh$ distribution is dominated by small BHs. 
In this section, we also evaluate how likely it is that a BH with a given $\mbh$ would undergo a merger, by comparing the $\mbh$ distribution of the merger events with the underlying BH mass function.
Figure~\ref{fig:BH-merger-mbh} shows the averaged (expected) rate for a BH from a given $\mbh$ bin to merge with another BH, with the orange, green and blue lines showing results at three different redshifts, $z=5$, $z=4$ and $z=3$ respectively. 
The $y$ axis is in units of merger rate ($\rm{Myr}^{-1}$) at a given redshift $z$, obtained by averaging the merger rate calculated in a 300 Myr bin centred on $z$, over the BHs in each $\mbh$ bin at that redshift.
For example, the blue solid line suggests that at $z=3$ a BH of mass $10^7 \msun$ would be expected to undergo about 1 merger per Gyr.
The solid lines include all BH mergers, while the dashed-dotted lines include only the non-seed BH mergers where both BHs have $\mbh > 2\times \msd$ at the time of the merger.
Including only the non-seed BH mergers reduces the expected merger rate by $0.3\sim0.5$ dex. 
The dashed-dotted lines drop off at $\mbh < 10^6 \msun$, as most of the BHs there are still at the seed mass.

The rates shown in Figure~\ref{fig:BH-merger-mbh} do not exhibit significant time evolution, with a similar trend of increasing merger rate with BH mass holding at all redshifts.
The $10^8 \msun$ BHs have an average expected merger rate about $10$ times higher than the $10^6 \msun$ BHs. 
This trend is mainly due to the fact that the most massive BHs reside in the most massive galaxies that have a large BH occupation number (as shown in Figure~\ref{fig:Occupation}), and therefore they have the largest possibility to encounter mergers.

\begin{figure}
\centering
  \includegraphics[width=0.9\columnwidth]{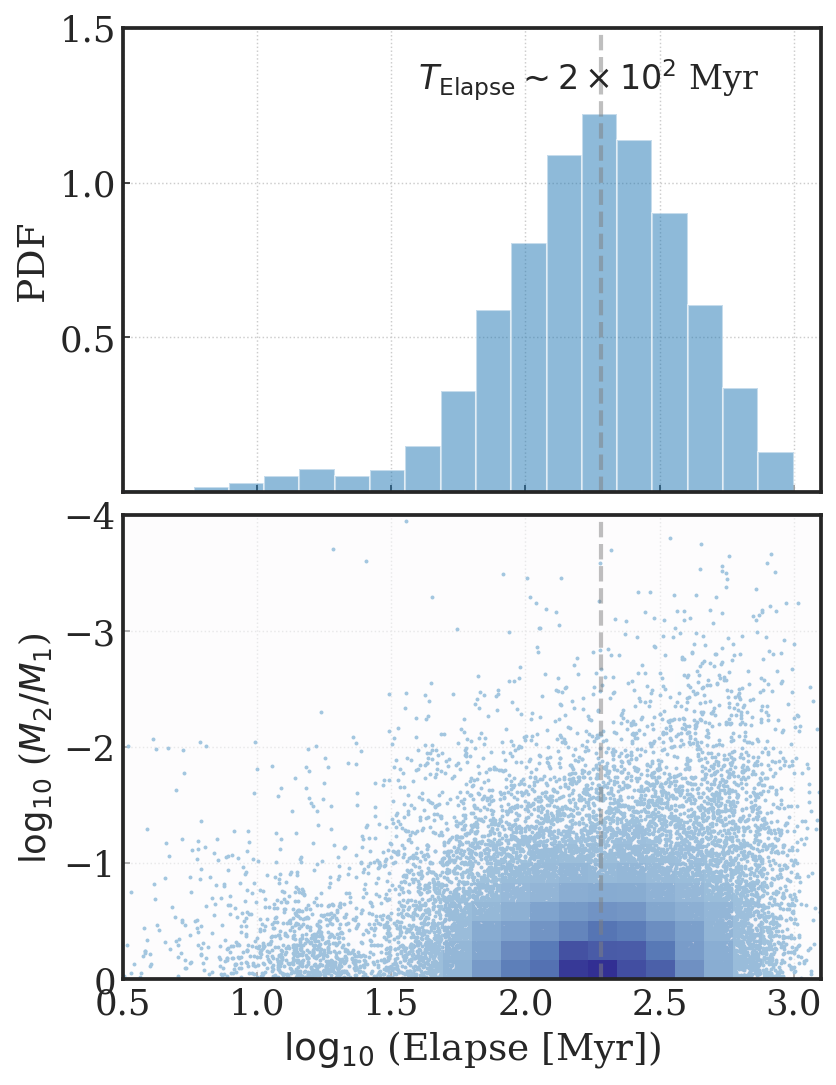}
  \caption{\textit{First panel}: Distribution of the elapse time (defined as the time between the first encounter of the two BHs and final merger time) for all the non-seed mass BH merger events. 
  \textit{Second panel}: Relationship between the elapse time and the mass ratio $q \equiv M_2/M_1$ of the BH binary.
  The vertical dashed lines in each panel mark the most probable elapse time, which is  $t_{\rm elapse} \sim 2 \times 10^2$ Myrs.}
  \label{fig:BH-merger-delay}
\end{figure}

\begin{figure*}
\centering
  \includegraphics[width=1.0\textwidth]{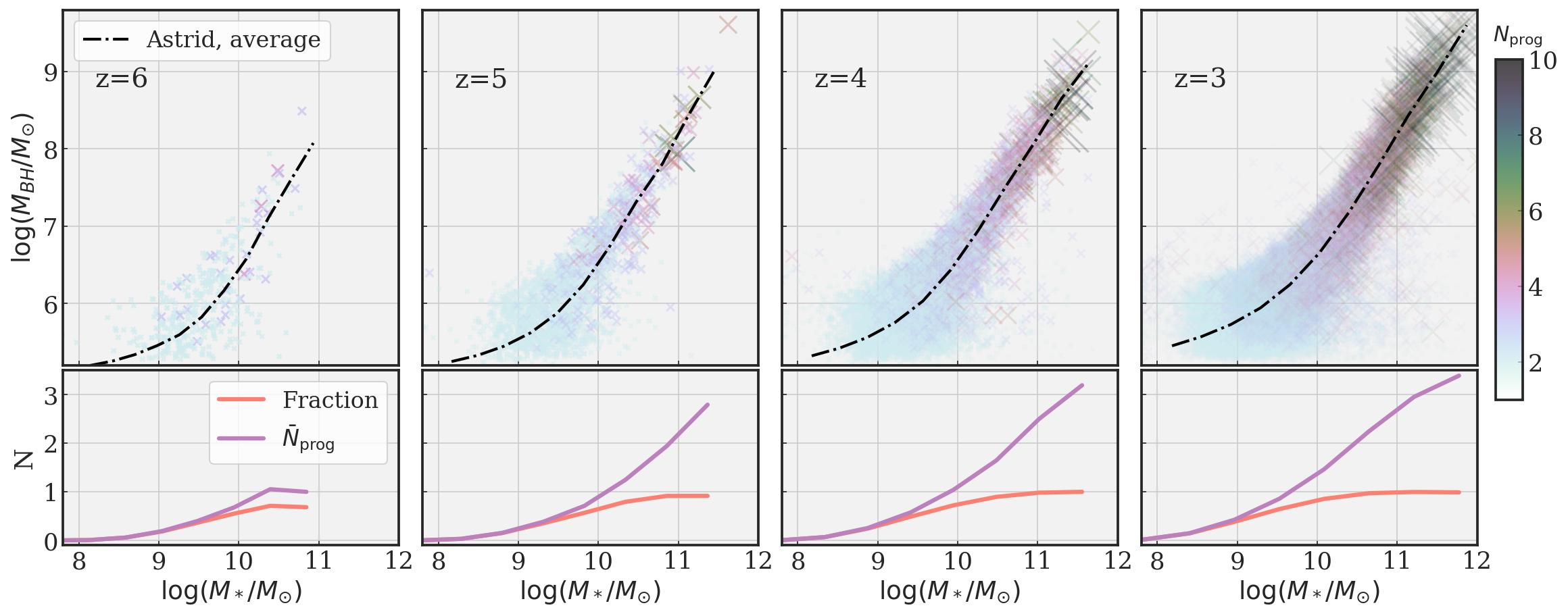}
  \caption{
  \textit{Upper panel:} The coloured crosses in each panel show the $\mbh$ - $\mstar$ relation for the BHs that have undergone mergers at each redshift. The colours and sizes of the crosses are scaled using the number of BH progenitors. The black dashed-dotted lines are the averaged $\mbh$ - $\mstar$ relation for that redshift as given in Figure~\ref{fig:scaling}. 
  \textit{Lower panel:} The purple curves show the averaged number of BH progenitors in each stellar mass bin. The red curves show the expected fraction of galaxies that contain at least 1 BH progenitor (that have undergone at least 1 BH merger event.)}
  \label{fig:Scaling-merging}
\end{figure*}

\subsubsection{Elapse time of BH mergers}

In this section, we investigate the timescale over which BH binaries spiral towards each other before a merger. 
This characterises the typical time scale for dynamical friction to take effect by dissipating the BH kinetic energy and making the binary system gravitationally bound.
For each of the merger events, we trace the BHs evolutionary histories and determine the first time the two BHs encounter each other (i.e., satisfy the distance criteria for BH mergers, $|\bf{\Delta r}|$ $< 2 \epsilon_{\rm g}$). We define the elapse time for a BH merger, $t_{\rm elapse}$, in \asterix as the time between the first encounter and the final merger.

The top panel of Figure~\ref{fig:BH-merger-delay} shows the overall 1D histogram of elapse times for all non-seed mass mergers. 
The bottom panel is a scatter plot of $t_{\rm elapse}$ and the mass ratio $M_2/M_1$ for the merger events.
We choose not to show the seed-mass population for two reasons: first, this population dominates the entire merger population, leading to a concentrated high-density region in the bottom panel which makes the relationship hard to see; second, the seed-mass BHs with boosted $M_{\rm dyn}$ can upset the relationship between dynamical friction time scale and the BH masses.

From the 1D distribution, we can see that the elapse time of BH mergers peaks at around $t_{\rm elapse} \sim 200$ Myrs.
This is a significant delay compared to the BH first encounter time. 
In fact, as will be shown in \cite{Chen2021b}, the resolved dynamical friction time is already more than the unresolved dynamical friction time. 
This indicates that by adding the sub-grid dynamical friction model, we can already account for more than half of the orbital decay due to dynamical friction directly within the simulation.

Analytical estimates \citep[e.g.][]{Binney2008} expect that the dynamical friction time is proportional to the velocity dispersion of the primary galaxy, and is in turn positively correlated with the mass of the primary BH; It is also inversely proportional to $M_2$ and should therefore result in a negative correlation between $M_2/M_1$.
Although we observe hints of this in the 2D distribution, it is very weak, implying that other environmental factors (such as the stellar density, morphology of the host galaxies, eccentricity of the BH binary orbit etc.) are also important to determine the actual elapse time for the BH mergers in the simulation.


\subsection{Statistics of the host galaxies for BH mergers}
\label{subsection: bh-merger-host}

In this section, we examine where the merging BHs reside with respect to the $\mbh$ - $\mstar$ relation.
In the upper panel of Figure~\ref{fig:Scaling-merging}, we identify all the BHs that have undergone at least 1 merger event and plot their host galaxy mass in the $\mbh$ - $\mstar$ plane. The cross markers are coloured and scaled by the number of the BH progenitors.
The black dashed-dotted lines are a repeat of those shown in Figure~\ref{fig:scaling}, giving the average $\mbh$ - $\mstar$ relation from $z=6$ to $z=3$ in \asterix.
It is interesting to see that the cross markers evenly populate the area around the mean $\mbh$ - $\mstar$ relation, indicating that the systems hosting BH mergers do not have a biased distribution compared to the overall population. 
We note that some other simulations that perform BH mergers without modelling dynamical friction also find a similar feature \citep[see][]{Degraf:2021-inprep}.

The lower panels of Figure~\ref{fig:Scaling-merging} show the statistics of BH mergers with respect to the stellar masses of host galaxies at each redshift. The red lines give the fraction of galaxies that have undergone at least one BH merger event in the past, and the purple lines give the averaged number of BH mergers that have occurred in the galaxies in each stellar mass bin.
We can see that more massive galaxies have a higher possibility to host BH mergers.
After $z=5$, the fraction of mergers saturates for large galaxies with $\mstar \sim 10^{11} \msun$, indicating that for galaxies of this size, all of them host BHs that have undergone mergers.
The average number of BH mergers increases with the stellar mass. 
This is because, as shown in Figure~\ref{fig:Occupation}, massive galaxies typically have high occupation numbers of BHs and experience a relatively large number of BH mergers. 

\section{Summary}
\label{section6:Conclusion}

In this work, we have analysed the evolution of the massive BH population in the large volume cosmological hydrodynamical simulation \asterix at $z \geq 3$.
\asterix uses a BH accretion and AGN feedback model evolved from that used in the \bluetides~simulation. In order to broadly capture the effects of different BH seeding mechanisms as well as BH growth below the resolution of our simulations, BHs are seeded with a range of BH seed masses. 
Each $\msd$ is stochastically drawn from an inverse power-law distribution ranging from $M_{\rm sd,min} = 10^4 \hmsun$ to $M_{\rm sd,max} = 3 \times 10^5 \hmsun$, allowing us to investigate the effects of seed mass on the subsequent BH growth.
We have replaced the artificial 'repositioning' of BHs to a local minimum with the implementation of dynamical friction from collisionless particles. Dynamical friction causes the BHs to sink into the central region of galaxies and leads to improved physical modelling of the BH mergers.
The explicit inclusion of dynamical friction and the large cosmological volume (a $250 \hmpc$ box) allows us to make a number of predictions for multi-messenger studies of BHs.

We have presented statistical properties of the BH population, such as their $\mbh$ and luminosity distributions spanning $z>6$ to $z=3$. 
The global evolution of the BH mass density and BH accretion rate is broadly consistent with current observational constraints. 
The BH mass function displays good agreement with observationally inferred results. 
We find that the BH luminosity $L_{\rm bol}$ (derived from the accretion rate) exhibits a positive correlation with $\mbh$, with the mean $L_{\rm bol}$-$\mbh$ relation decreasing with time over the redshift range from $z=7$ to $z=3$. 
During this time the relation also becomes progressively flattened at the more massive $\mbh$ range, consistent with the fact that for the most massive BHs, on average, the Eddington accretion ratio $\lambda_{\rm Edd}$ decreases with redshift. 
This effect can be interpreted as the result of AGN feedback and decreased cold gas fraction at lower redshift.

We compare the BH luminosity function to multiple observational constraints in the hard X-ray band.
Our predicted X-ray luminosity functions agree well with observations at $\log \Lx > 43$. The intrinsic AGN luminosity function has a somewhat steeper slope than observations at the faint end with $\log \Lx < 43$.
We estimate the extent of galactic scale obscuration in each source by computing the hydrogen column density $\NH$ surrounding the AGN and accounting for angular variations corresponding to different sightlines.
The galactic obscuration contributed by ISM gas is large ($\NH \sim 10^{23-24}$ cm$^{-2}$) for high-redshift AGN with $z>6$ and decreases to $\NH \sim 10^{22-23}$ cm$^{-2}$ at lower redshift $z=3$. 
We find that at all redshifts the galactic column densities increase with AGN luminosity.
Interestingly, this trend is opposite to what is generally found in observations, where the faint AGN population is typically more obscured than the bright one.
At $z=3$, the galactic component of the obscuration falls short at reproducing the large column densities constrained from X-ray observations. This implies that while a galactic gas component can provide substantial obscuration from $z = 6 \sim 3$, there is a significant contribution from a nuclear torus and this component is crucial for explaining the large amount of obscuration at the faint end of the AGN LFs.

The UV luminosity of AGN from \asterix compares favourably with the available observations, particularly when considering magnitudes $M_{\rm UV} < -23$. 
At $M_{\rm UV} > -23$, however, most of the AGN has a UV luminosity fainter than their host galaxy.
Because of this, and the poorly constrained gas to dust conversion at these redshifts, we find that the number density of the faint UV AGN population is hard to predict to within a factor of at least $10$.

We investigate the effects of the BH seed mass on the BH growth by tracing BHs seeded at $z>10$ down to $z=3$.
We find that at $z=3$, BHs with $\msd > 10^5 \msun$ carry virtually no imprint of their initial $\msd$. However, a large fraction of BHs with the smallest $\msd$ has not grown significantly, more than half of them have not doubled their mass at $z=3$.
The trace of the original $\msd$ is erased with time evolution as a result of the substantial gas accretion which governs the growth of massive BHs.
We decompose the BH growth into the two different channels of accretion and mergers, showing that BH accretion is the dominant contributor to BH growth, while BH mergers play a subdominant role.


The scaling relations, $\mbh-\mstar$ and $\dot{M}_{\rm BH}$-SFR, in \asterix reveal correlations between the growth of BHs and their host galaxies which are broadly consistent with observations (for $\mstar \geq 10^{10} \msun$).
The $\mbh-\mstar$ relations at $z=3-6$ have a slightly lower normalization than the observed relations at $z=0$.
We have also found that this relation flattens and the scatter increases at $M_{\rm BH} \leq 10^{6} \msun$.

The galactic column density contributions to the AGN have a weakly positive correlation with the host galaxy stellar mass (broadly consistent with what was inferred from observations at $z=2-4$). We find a large scatter in the column density, as expected by the large angular variations in the gas surrounding the BH and the effect of AGN feedback and outflows \citep[see also][]{Ni2020}.

With the improved modelling of the BH dynamical friction (and the removal of `repositioning'), we were able to examine BH occupation statistics for the galaxies in \asterix. 
Quantitatively, we found that galaxies with $\mstar \geq 10^{11} \msun$ host an average of $>10$ BHs.
However, most of these are close to the seed mass and even in the large galaxies the average occupation number of luminous BHs ($\Lx > 10^{43}$ erg s$^{-1}$) remains only $1 - 2$. 

We analyzed the global statistics of BH mergers in \asterix.
There are in total 450 thousand BH mergers in \asterix at $z>3$.
Most of the mergers involve seed mass BHs and only a few thousands occur for bright AGNs ($\Lx > 10^{43}$ erg s$^{-1}$) that have at least one detectable electromagnetic X-ray counterpart. 
When considering the number of mergers scaled by the BH mass function of the whole BH population, we find that the massive BHs ($> 10^{7} \msun$, which likely to have detectable EM signatures) experience the most frequent mergers. 
These are the objects that reside in more massive galaxies, with high BH occupation numbers and typically experience a relatively large number of BH mergers.

The effects of directly modelling the BH dynamical friction predicts a typical elapse time for a merger event (the time between the first passage of the BH pair and their final merger) of about $2\times10^2$ Myrs.
This is an important effect that leads to a significant delay from the time of the first encounter (when for example, with repositioning to the local potential minimum, the BHs would have already `merged').
Interestingly, we have shown that the galaxies that host BH mergers trace out the same $\mbh-\mstar$ relation as the full BH-galaxy population.  
Massive galaxies host the majority of BH mergers as they have a high occupation number of BHs. 

In this paper, we have provided a broad overview of the main results for the BH growth and evolution at $z>3$ in the \asterix simulation.
In particular, we have presented primary results on BH statistics and the relationship between BHs with their host galaxies. 
We have made predictions for the BH merger rates and BH occupation fractions enabled by our self-consistent modelling of BH dynamical friction. 
We reserve to shortly upcoming works more detailed investigation and several further applications, such as making mock X-ray catalogues and quasar spectra and using detailed galaxy and AGN SEDs to disentangle the galaxy and AGN contributions.
We plan to create BH merger catalogues with associated predictions of electromagnetic counterparts (for the AGN and the host galaxy). 
We will make detailed predictions for the detectability of these events by LISA (considering also, as part of postprocessing, other BH binary evolution processes and orbit eccentricities).


\section*{Acknowledgements}
\asterix is carried out on the Frontera facility at the Texas Advanced Computing Center. The simulation is named in honour of Astrid Lindgren, the author of Pippi Longstocking. 
We thank Aswin Vijayan, Steve Wilkins, Mislav Balokovic for helpful discussions. 
SB was supported by NSF grant AST-1817256.
TDM and RACC acknowledge funding from the NSF AI Institute: Physics of the Future, NSF PHY-2020295, NASA ATP NNX17AK56G, and NASA ATP 80NSSC18K101. TDM acknowledges additional support from  NSF ACI-1614853, NSF AST-1616168, NASA ATP 19-ATP19-0084, and NASA ATP 80NSSC20K0519, 
and RACC from NSF AST-1909193.
MT is supported by an NSF Astronomy and Astrophysics Postdoctoral Fellowship under award AST-2001810. 

\section*{Data Availability}
The code to reproduce the simulation is available at \url{https://github.com/MP-Gadget/MP-Gadget} in the \texttt{asterix} branch. 
Halo catalogues and snapshot particle tables are available on reasonable request to the authors.
The full BH catalogues at $z>3$ will be available online shortly.

\bibliographystyle{mnras}
\bibliography{bib.bib}

\end{document}